# Inactive Overhang in Silicon Anodes

Aidin I. O'Brien,[1] Stephen E. Trask,[1] Devashish Salpekar,[1] Seoung-Bum Son,[1] Alison R. Dunlop,[1] Gabriel M. Veith,[2] Wenquan Lu,[1] Brian J. Ingram,[1] Daniel P. Abraham,[1] Andrew N. Jansen,[1] Marco-Tulio F. Rodrigues[1,z]

[1] Chemical Sciences and Engineering Division, Argonne National Laboratory, Lemont, IL, USA

[2] Chemical Sciences Division, Oak Ridge National Laboratory, Oak Ridge, TN, USA

[z]**Contact:** marco@anl.gov

**Abstract**

Li-ion batteries contain excess anode area to improve manufacturability and prevent Li plating. These overhang areas in graphite electrodes are active but experience decreased Li$^+$ flux during cycling. Over time, the overhang and the anode portions directly opposite to the cathode can exchange Li$^+$, driven by differences in local electrical potential across the electrode, which artificially inflates or decreases the measured cell capacity. Here, we show that lithiation of the overhang is less likely to happen in silicon anodes paired with layered oxide cathodes. The large voltage hysteresis of silicon creates a lower driving force for Li$^+$ exchange as lithium ions transit into the overhang, rendering this exchange highly inefficient. For crystalline Si particles, Li$^+$ storage at the overhang is prohibitive, because the low potential required for the initial lithiation can act as thermodynamic barrier for this exchange. We use micro-Raman spectroscopy to demonstrate that crystalline Si particles at the overhang are never lithiated even after cell storage at 45 °C for four months. Since the anode overhang can affect the forecasting of cell life, cells using silicon anodes may require different methodologies for life estimation compared to those used for traditional graphite-based Li-ion batteries.

**Introduction**

Li-ion batteries (LIBs) are commonly manufactured with an excess area of the negative electrode (NE, or anode) coating. These "overhangs" increase the tolerance to manufacturing imperfections by guaranteeing that there are always NE areas juxtaposed to the positive electrode (PE, or cathode),[1,2] decreasing cell-to-cell variability and helping prevent Li deposition at the NE edges.[3] The anode overhang (AOH) typically comprises a patch of coating that is ~1-mm wide throughout the length of the electrode,[1] effectively increasing the area of the NE by ~3% or more, depending on the cell format. In practice, the total area contributed by the AOH to the anode can be significantly larger, as the outermost portions of the double-sided NE coating will lack a PE counterpart in certain commercial cells.[4] A consequence of this design feature is that the total coated area of the anode may exceed that of the cathode by as much as ~12%.[5] These AOH regions can be classified according to their relative distances to the cathode coating, into near-overhang (the millimeter-long strips) and far-overhang (the outer portions of the NE).[6]

The impact of the overhang on cell behavior for graphite (Gr) anodes has been discussed in prior works.[1,2,5-13] During cycling, most of the $Li^+$ flux into and out of the anode will occur in the portions directly opposite to the cathode coating, which defines what is known as the "active area" (AA) of the NE. Although direct lithiation of the overhang area is possible, the longer distances between the cathode and the AOH will limit the extent of this process,[2] establishing a gradient in the lithium content between the AA and the AOH. This difference in $Li^+$ concentration creates a driving force for homogenization through the slow diffusion of lithium ions through the active (Gr) particles. Another option for equalization of the $Li^+$ content between AA and AOH is migration: the plateaus in the voltage profile of Gr (Figure 1a) create a gap in electrical potential as a function of the degree of lithiation, driving homogenization for as long as this gap persists.[1,2]

In this case, faster ionic transport is assisted by the electrolyte,[1, 2, 5, 9] while electrons travel through the conductive network of the NE. The end result is that, over time, lithium ions will accumulate at the overhang, as it continuously tries matching the Li$^+$ content of the active area. In certain cases, a lithiated AOH can also act as the provider of Li$^+$, donating lithium ions to the AA when the cell is at lower states of charge (SOCs). Either way, these exchanges are slow, extending from hours to weeks for the near-AOH or many months for the far-AOH.[5, 6] Higher temperatures have been shown to accelerate these equalization processes.[1, 5]

Given the time scales over which equalization processes will persist, the overhang can have meaningful effects on measurements of capacity,[1, 7] coulombic efficiency (refs. [1, 9, 14]) and self-discharge.[11, 12] Notably, these slow Li$^+$ exchanges can artificially deplete or replenish the measurable Li$^+$ inventory in the cell by several percent, leading to instances of capacity gain during aging or to a systemic overestimation of growth rates for the solid electrolyte interphase (SEI). The AOH acts as a reversible sink or source of lithium ions with kinetic limitations,[1, 7] carrying the signature of a "memory" of the prior history of the cell.[9]

Since the overhang can distort capacity measurements over several months, it can have significant effects on the perceived trajectories of aging. Calendar aging of commercial Gr-based cells is dominated by reduction reactions at the NE, causing SEI growth that leads to capacity fade. This fade has been widely reported to taper down after reaching a maximum rate early during testing, in accordance with the idea that growth of the passivation layer becomes self-limiting according to a $\sqrt{t}$ dependence, where $t$ is the time for which cells have been aged. Nevertheless, recent works have questioned whether this ubiquitous trend is simply an artifact introduced by Li$^+$ outflow into the overhang.[2, 10, 15-17] Commercial cells leave manufacturing plants at low SOCs to minimize calendar aging prior to deployment. The AOH will slowly receive Li$^+$ from the AA for as long as cells remain idle, causing the overhang to inch towards this same low $Li^+$ content. When cells are finally charged and monitored during calendar aging studies, the higher SOC will prompt additional lithium transport towards the overhang, adding to the measurable decrease in $Li^+$ inventory caused by SEI growth. This months-long overhang equalization will eventually subside once both AA and AOH reached the same electrical potentials, and could explain the initially faster capacity fade.

Additional evidence of the influence of the overhang over initial trajectories of aging was presented by Lewerenz et al.,[10] who showed that calendar aging at high states of charge led to linear rates of capacity decay when cells had been initially conditioned at 99% SOC for several months. Interestingly, storage of these pre-conditioned cells at low SOCs led to capacity gains for several months due to recovery of $Li^+$ from the AOH, leading to aging profiles that resembled "inverse" $\sqrt{t}$ traces.[10] Regardless of the actual functional form describing SEI growth, there is substantial evidence that a complete physical description of aging trends must consider the reversible but slow $Li^+$ exchanges between the active area and the overhang. This may be especially relevant in the context of life forecasting, when attempts to project future cell performance are made after sampling data for a limited period. In this case, initial overhang activity could be propagated into the future at the expense of relevant aging mechanisms, either leading to inaccuracies in estimates or constraining the required size of the dataset by how long it is needed to properly describe the AOH equalization.

Our team is particularly interested in life forecasting for cells containing Si-rich anodes. High-energy cells based on silicon can exhibit calendar lives that are notably lower than that of traditional graphite-based systems,[18] and methodologies that allow the inference of life from relatively short experiments are extremely important to accelerate discovery of mitigation strategies.[19] Although there exists a substantial amount of work on life forecasting for commercial cells (with ≤ 5 wt% Si), these typically do not include an explicit description of capacity exchanges with the overhang. Hence, it is possible that direct transferability of such models to Si-dominant systems may depend on whether overhang equalization proceeds according to the same tenets and timescales as for graphite anodes. Here, we show that overhang activity in Si particles can differ greatly from that observed for traditional Gr anodes.

The crux of our hypothesis relates to the hysteresis and the shape of the voltage profile of Si electrodes (Figure 1), and to the energetics of overhang equalization. In Gr, operation at slow rates will lead to only a small gap between the voltage profiles during lithiation and delithiation (Figure 1b). This finite separation will persist even after extended relaxation, due to disorder of intercalated $Li^+$ across the Gr planes.[20] Nevertheless, even with this persistent gap, the potential for the lowest plateau during delithiation (stage 1, at ~97 mV) still remains lower than that for the stage 2 plateau for lithiation (~113 mV). Consequently, when the AA is at stage 1 and the AOH has a much lower $Li^+$ content, a driving force for equalization can be maintained until the overhang starts forming $LiC_6$ (Figure 2, top). This homogenization, when the cell is at rest, must occur without external input of energy, and the small hysteresis of Gr guarantees that this process remains possible even when the energy penalty associated with $Li^+$ extraction/insertion from the lattice (i.e., the hysteresis) is considered. This may not be the case for Si.

Consider the voltage profiles for a Gr-free electrode containing amorphous Si (a-Si) particles, which presents monotonic changes in potential as a function of lithium content for both lithiation and delithiation (Figure 1c). A ~75 mV rebound is observed when the half-cell is allowed to rest after reaching 50 mV. Regardless of how slow the cell is cycled or how long the cell is allowed to rest, a sizable voltage gap will always exist (Figure 1d),[21-23] and thus changes in lithium content will require very different potentials depending on the direction of change. It has been suggested that this hysteresis could arise as a consequence of complex mechanical behavior involving the SEI under large strains.[21] We hypothesize that the hysteresis will decrease the driving force for overhang equalization to the point that it may not happen in Si electrodes. Considering Figure 1d, due to the energy penalty required for delithiation, electrons extracted from the AA after lithiation to 50 mV would have the reducing power equivalent to ~200 mV, only sufficient to bring the AOH to much lower states of lithiation than that of the active area. Achieving higher levels of equalization would violate energy conservation. Furthermore, the lack of plateaus in the voltage profile causes a progressive decrease in driving force for equalization, as $Li^+$ transport to the overhang decreases the lithium content of the neighboring AA. Si anodes with this lower Li content would only delithiate at even higher potentials, further lowering the reducing power available to lithiate the overhang (Figure 2, bottom). An even more prohibitive case occurs for Gr-free electrodes containing crystalline Si (c-Si), which must reach ~110 mV before lithiation can initiate (Figure 1e,f).[24] The large hysteresis of Si may cause it to be virtually impossible for the delithiation of the AA to release sufficient energy to drive lithiation into the AOH. Hence, while overhang utilization in Gr is limited by kinetics, an additional thermodynamic barrier imposed by hysteresis may exist in Si, further restricting $Li^+$ homogenization within the electrode.

In the present work, we use Raman spectroscopy to demonstrate that the anode overhang in crystalline Si electrodes is never lithiated even after full-cells are rested at high SOCs and elevated temperatures for extended periods. Since c-Si is irreversibly amorphized the very first time it lithiates, and a- and c-Si have distinct Raman-active modes, this technique allows us to "see back in time" and determine *ex situ* if regions had ever been lithiated during the cell life. Interestingly, we also show that, despite the inactivity of Si at the overhang, electrons are still able to permeate through the conducting network at the AOH, causing those areas to contribute to aging via SEI growth. We finally extrapolate our results and discuss how overhang equalization is expected to proceed in other systems, such as silicon-graphite blends. Given the differences we report between Gr and Si, life models may only be able to universally describe cell aging if the behavior of the overhang is explicitly accounted for.

*Experimental*

***Electrode composition and cell assembly.*** All electrodes used in this study were fabricated at Argonne's Cell Analysis, Modeling and Prototyping (CAMP) Facility. Nanoparticles (~250 nm) of c-Si were obtained from milling of Si boules.[25] Crystalline silicon electrodes comprised 79 wt% of c-Si, 9.87 wt% C-45, 9.88 wt% of an acrylate binder (Battbond 290S, Blue Ocean and Black Stone), 0.5 wt% of single-walled carbon nanotubes (Tuball) and 0.75 wt% carboxymethyl cellulose (CMC). The CMC was present in the nanotube dispersion. Total coating loading was 1.19 mg/cm$^2$ (~2.5 mAh/cm$^2$, 0.05 – 0.7 V vs. Li/Li$^+$) and coating thickness was 17 μm. The electrode was not calendared and had an estimated porosity of ~53%. The c-Si electrode was heat treated at 350 °C for 1 hour prior to use; we have observed that this treatment improves cyclability of the electrode. The positive electrode used for all Si-containing full-cells comprised 96 wt% of LiNi$_{0.8}$Mn$_{0.1}$Co$_{0.1}$O$_2$ (NMC811, Targray), 2 wt% of C-45 carbon additive (Timcal) and 2 wt% of polyvinylidene fluoride (PVDF, 5130, Solvay). The electrode had 75 μm of coating thickness (34.3% porosity after calendering) and a total coating loading of 22.02 mg/cm$^2$ (~3.95 mAh/cm$^2$, 3.0 – 4.2 V vs. Li/Li$^+$).

Some experiments involved Gr electrodes tested in full-cells. In that case, the negative electrode consisted of 91.83 wt% of Gr (SLC1506T, Superior Graphite), 2 wt% C-45 carbon, 6 wt% PVDF (9300, Kureha) and 0.17 wt% oxalic acid. The electrodes presented a coating loading of 9.38 mg/cm$^2$ (~2.84 mAh/cm$^2$) and a calendered coating thickness of 70 μm (38.2% porosity). The positive electrode used in Gr full-cells had similar composition as the one listed above, but with a 53-μm coating (35.2% calendered porosity) and 15.35 mg/cm$^2$ total coating loading (~2.7 mAh/cm$^2$, 3.0 – 4.2 V vs. Li/Li$^+$). Information for other electrodes used in this study is provided in Table S1 of the *Supplementary Material*. Gr and NMC811s were dried overnight under dynamic vacuum at 120 °C prior to use. Cathodes were coated on 20 μm aluminum and anodes on 10 μm copper foils, both battery-grade. The electrode coating was carried out through a roll-to-roll process using an A-PRO coater, equipped with a 25- mm-wide reverse comma bar and a two-zone-temperature drying oven (1 m per zone).

Electrochemical characterization was performed in 2032-type coin cells. Two full-cell configurations were tested, including cathode / anode pairs with diameters of either 10/14 or 14/15 mm. These configurations resulted in cells containing overhangs with areas equal to 96% or 15% of the active area, respectively, and we will use these numbers to differentiate these cells throughout the manuscript. The separator was a 16 mm disc of Celgard 2500. The electrolyte was a 1.2M solution of LiPF$_6$ in 3:7 wt:wt of ethylene carbonate and ethyl methyl carbonate (Tomiyama), with 3 wt% fluoroethylene carbonate (FEC, Solvay). Each cell contained 40 μL of electrolyte, which is >4x the total pore volume of cell components. Cell assembly and teardown were performed in a dry room (dew point < -45 °F).

*Electrochemical testing.* All cycling was performed at 30 °C on a Maccor Series 4100 cycler. Calendar aging experiments involved charging cells to the desired voltage, disconnecting them from the cycler and storing them at 45 °C for the designated duration. Elevated temperatures are expected to accelerate the AOH equalization.[1, 5, 9] All full-cells went through a wetting process prior to cycling, consisting of a 4-hour rest in the environmental chamber, a charge to 1.5 V followed by a voltage hold until 15 minutes were completed and then an additional 4-hour rest. Formation cycles for c-Si full-cells involved charges to 3.9, 3.85 and 3.9 V, all using 3 V as discharge cutoff; the charge cutoffs were devised to supply sufficient $Li^+$ to compensate for initial formation of the solid electrolyte interphase (SEI). Additional cycling was performed between 3 and 3.9 V, effectively cycling the anode between ~0.05 and ~0.6 V vs. $Li/Li^+$. Lithiating the Si beyond 0.05 V led to worsened performance due to mechanical effects. Typically, formation cycles were performed at C/10. In specific experiments, we evaluated whether the AOH utilization depended on the C-rate of formation cycles, and the utilized protocols are described in the text when relevant. Graphite-based full-cells were always tested between 3 and 4.1 V.

To probe AOH equalization during cycling, c-Si and Gr full-cells were tested at a slow C/10 rate for a total of 30 cycles, there being for formation (as described above) and 27 for aging. For assessing the role of the AOH during calendar aging, several full-cells went through formation cycles and were then charged to the desired voltage cutoff (3.6 or 3.9 V for c-Si, or 4.1 V for Gr). Cells were then stored at 45 °C for the desired period (0.5, 1, 2.5 or 4 months), after which they underwent a discharge and a full cycle, all at C/10.

***Raman spectroscopy.*** Spectra were acquired using a Renishaw inVia Raman microscope located inside an argon-filled glovebox, with 785 nm laser excitation (100 mW). Optical images of the samples were initially taken with a 5x objective in the "montage" mode, resulting in a collage of micrographs of the entire surface of the electrode (Figure 3a). We then switched to a 20x objective without disturbing the sample stage, so that the 2D coordinates of zoomed-in images could be correlated with that of the collage of the entire electrode. The imprint left by the positive electrode was generally visible in the anodes after cell disassembly, indicating the approximate position of the overhang. Optical images around regions of interest surrounding this boundary were collected (Figure 3b), and a series of Raman spectra were recorded along a path traversing this boundary, typically with a 40-70 µm gap between adjacent measured areas. In most samples, spectra were collected from four different regions at different sides of the electrode, so that assessment of AOH utilization becomes less dependent on assumptions made about the placement of the active area. Spectra were collected using a 20x objective and with a single acquisition of 45-60 s, with power limited to 10% of the total laser power. A total of over 200 spectra were typically collected for each sample.

Post processing of the acquired spectral data involved truncating spectra to within the ~200-800 $cm^{-1}$ range. A linear baseline was subtracted based on the spectral intensity at 560-600 $cm^{-1}$; this region is devoid of active modes and was selected to bring all spectra to a similar intensity scale to facilitate batch processing. We did not attempt to subtract a "true" baseline from the spectra since the absolute band intensities were unimportant for our analysis and because Si bands at low wavenumbers made it difficult to avoid possibly distorting the spectra.

We used Python algorithms for data analysis and presentation. We used the *opencv* package to create an interface that would let us select specific points along the *x* and *y* axes of the full-electrode collages (Figure 3a). The axes lines were one-pixel tall and wide, enabling accurate picking of points. The algorithm would then take the coordinate values of the selected points as input, and use them to calibrate a correlation between pixel location in the image and (x,y) coordinates in the axes scales. Once that calibration was established, the algorithm would take (x,y) coordinates for the centroid of the circle describing the active area, drawing that over the collage; this process was iterative, and was repeated until the drawn yellow circle matched the cathode imprints visible in the electrode. Afterwards, the algorithm would take the edge coordinates for the zoomed-in areas imaged with the 20x lens and use that to draw red rectangles over the collage, indicating the locations where the overhang had been probed. Finally, the algorithm would load a high-resolution version of the collage (without axes), draw the relevant shapes over it and export the final figure (Figure 3c). The aspect ratio between the low- and high-resolution versions of the collage were different, causing a small distortion in the latter image. The algorithm accounted for this by readjusting the calibration and drawing the circle describing the overhang boundary as an ellipse with low eccentricity, instead.

A second algorithm was created to analyze and then represent the spectral data over micrographs of each probed region. The algorithm would attempt to fit Gaussian (for the a-Si) and Voigt (for the c-Si) curves to all spectra. The allowable range for the centroid of each of these curves was selected as appropriate ($455 - 463$ $cm^{-1}$ range for amorphous and $512 - 525$ $cm^{-1}$ for crystalline Si). The ratio between the fitted intensities was used as a first attempt to define whether the specific band was present in each dataset. The algorithm would then use a procedure similar to that described above to correlate pixels and (x,y) coordinates in the optical images (collected with the 20x objective). Dots were then drawn over the micrograph at the coordinates where spectra were collected (Figure 3d). These dots were color-coded to facilitate the visualization of the spatial distribution of silicon phases: black when only a-Si was detected, pink for regions were both amorphous and crystalline Si are observed (regardless of the relative intensity of each band), and red when only c-Si was present (Figure 3e). Although the algorithm would make an initial decision about this color assignment, all spectra were visually inspected and color tags were manually adjusted as needed. Finally, the algorithm used the centroid provided for the active area to draw an elliptical arc over a high-resolution version of the micrographs, to indicate the boundary between the active area and the AOH. An example of the resulting image is shown in Figure 3d. Occasionally, small mismatches between the arc and the cathode imprint would be observed in the zoomed-in micrographs, in which case we would perform additional adjustments to the centroid of the active area until a satisfactory placement was obtained in all images.

*Results and Discussion*

*Inferring overhang utilization using Raman spectroscopy.* Overhang equalization is a relatively slow process that can persist for many months. To evaluate whether Si particles at the AOH can become lithiated, we charged NMC811 vs. c-Si cells to 3.9 V (wherein the negative electrode would reach ~50 mV vs. Li/Li$^+$) and stored them at 45 °C for periods ranging from 0.5 to 4 months. We used Raman spectroscopy to analyze several of these cells and in all cases found the same result: Si at the overhang remained crystalline, indicating that it had never been lithiated. An example is shown in Figure 4, for a cell with 96% overhang aged for 4 months. Additional examples are provided as *Supplementary Material*, showing similar conclusions for cells aged for different periods (Figures S1,2) and with smaller AOH areas (Figures S3,4). Generally, spectra showing only the c-Si band were also detected within the final 50-100 µm of the active area surrounding the boundary with the AOH. Pink markers (denoting the detection of both a- and c-Si) also tended to become more prevalent closer to this boundary. Both these observations are in agreement with theoretical analyses, which suggest that the presence of an overhang will cause the outer portions of the AA to experience milder lithiation currents.[3, 26] "Mud cracks" in the coating are also only visible within the inner portions of the AA (Figure S15), providing additional visual indication of the areas where Si will preferentially cycle and endure severe stresses.

Figure 4 and related experiments support our hypothesis that, in c-Si electrodes, the concerted SOC homogenization required to propagate charge through the overhang is quite inefficient, constraining the anode activity to the active area. Although these tests indicated that lithiation of c-Si at the AOH during rest is negligible, it is still possible that Li$^+$ could be driven into Si particles in these outer regions of the electrodes during sufficiently slow cycling. All calendar-aged cells initially experienced three full cycles at C/10, so additional cells were made and underwent three formation cycles at even slower rates: C/25, C/50 and C/10 followed by a voltage hold at 3.9 V for 24 hours (when currents would drop as low as ~C/400). Regardless of how cells were cycled, the outcome was the same: significant Si lithiation appeared to proceed only within the active area (Figure 5 and Figures S5-8). Identical observations were found after continuously cycling cells at C/10 for nearly a month (Figures S9,10). Note that small variations in the coating loading of the electrodes can affect the extent of lithiation of silicon particles, contributing to the amount of residual c-Si observed within the active area (see Figures 5a-b, for example, where more *red* spots are visible within the AA even at slower cycling rates).

Now that we have provided evidence that c-Si indeed remains mostly inactive at the overhang, we explore additional nuances of this behavior.

***Consequences for cycle aging.*** In traditional Gr-based cells, Li$^+$ can slowly accumulate at the AOH. This Li$^+$ can remain kinetically trapped at the overhang and may not fully return to the positive electrode when the cell is discharged, creating the appearance of increased irreversible capacity fade (or gain, if the overhang initially has a higher lithium content than the active area).[1, 9, 10] Naturally, the larger the size of the AOH relative to the active area, the larger are these losses. Figure 6a shows the discharge capacities of NMC811-Gr cells (both 15% and 96% overhangs) as they experience 30 cycles at C/10; we chose this slow rate to favor AOH equalization. Capacities are normalized by the total weight of carbon materials (Gr + C-45) at the negative electrode within the active area. Cells with a larger AOH exhibit significantly lower discharge capacities already during the first cycle (Figure 6a), suggesting that parts of the overhang can already be accessed during the initial anode lithiation. Indeed, the initial coulombic efficiency (CE) is visibly lower when cells have a 96% overhang (Figure 6b). During cycling, capacity faded faster when the overhang was larger (Figure 6a), and the CE exhibited both a slower initial rise and lower overall values (Figure 6c). Very clearly, excess AOH area is detrimental to the performance of Gr cells right from the outset, in agreement with prior reports. [4, 8, 12, 13]

The cycling behavior of c-Si cells tested under the same conditions is strikingly different (Figures 6d-f). Although the gap in initial discharge capacity is still present, it is significantly more modest than the observed for Gr cells (Figures 6a). Widening of this gap is barely perceptible as the cells age. Overhang size also had lower effects on the initial CEs (Figure 6e), and both types of cells presented similar CEs throughout (Figure 6f).

The evidence from Raman measurements that c-Si at the AOH remains inactive (Figures S9,10) is corroborated by the very similar rate of fade between cells with the two designs. Nevertheless, the statistically significant gap in initial CEs is notable and, in principle, contradicts this view of an inoperant overhang. This conundrum is solved once we consider that the conductive network at the AOH can still engage in redox processes, even if the c-Si particles cannot be lithiated. The persistence of crystallinity at the overhang suggests that the AOH will always experience potentials above ~110 mV vs. Li/Li$^+$ (as lithiation and amorphization occurs at this potential). Cycling c-Si electrodes vs. Li metal between 150 and 700 mV will result in ~130 mAh/g$_{Si+C}$ of reversible capacity (Figure S12), roughly ~5% of the total capacity available in the electrode. We attribute this capacity to the C-45 and carbon nanotubes present in the NE, both of which exhibit highly sloped voltage profiles. Since these materials can store Li$^+$ at higher potentials than c-Si, it is likely that the carbons at the AOH become partially lithiated, extending the effective area of the electrodes where SEI can grow. Over time, this extended area can lead to additional levels of cell aging, even though Si particles at the overhang themselves tend to remain virtually inactive. We discuss this effect of time in more detail below.

***Consequences for calendar aging.*** An example of calendar aging test for a c-Si cell is shown in Figure 7a. Cells were first cycled three times during formation and then charged to the desired voltage prior to storage at 45 °C. Once the designated time had elapsed, cells were discharged and then underwent a final full cycle. The *irreversible* self-discharge during storage was quantified as the difference in discharge capacity between cycles 3 and 5. The *reversible* self-discharge was obtained from the difference between charge and discharge capacities in cycle 4, minus the irreversible losses. In both cases, values were normalized by the discharge capacity of cycle 3. While irreversible losses are generally a consequence of SEI growth, processes such as electrolyte oxidation at the positive electrode and the occurrence of redox shuttles can both decrease the available capacity (and cell voltage) during storage without permanently trapping electrons.[27-29] The decrease in cell potential as a function of total aging time was found to depend strongly on the size of the AOH (Figure 7b), with a faster decay observed for larger overhangs.

The irreversible self-discharge is shown in Figure 7c. Aging proceeded significantly faster for cells with 96% AOH, leading to the loss of ~38% of cell capacity after 4 months at 3.9 V and 45 °C (blue data points). In contrast, cells with 15% AOH only lost ~21% under the same conditions (orange markers). For comparison, Gr cells with 96% overhang and stored after being charged to 4.1 V lost ~60% of their initial capacity (light blue). Just as in the cycling experiments of Figure 6e, a larger c-Si overhang seemed to favor more capacity fade without being quite as extreme as Gr, a system in which the active material at the AOH can also actively and reversibly store $Li^+$. These findings agree with the view that, although Si particles are dormant at the AOH, the entire surface area of the conductive network within the electrode (both within the active area and overhang) can contribute to SEI growth. This is further supported by Figure 7e, which normalizes self-discharge by the total area of the anode. Inspection of the chart suggests that capacity losses per area of the anode are similar, highlighting the role of the overhang on cell aging. SEI growth at the overhang has also been suggested elsewhere.[8, 12] Additionally, several c-Si cells were aged at the lower 3.6 V (after the anode being lithiated to ~200 mV vs. $Li/Li^+$). Surprisingly, we found that the irreversible self-discharge was very similar to the observed for cells stored at 3.9 V (Figure 7c). In traditional Gr cells, aging generally correlates with how low the anode potential is.[30, 31] Our results seem to indicate that such dependency could be weaker for Si cells.

The reversible self-discharge accrued during storage at 45 °C is summarized in Figure 7d, and values were also found to be larger at increasing overhang sizes. Interestingly, similar values are observed for Si and Gr cells, despite the difference in aging voltages. We posit that this observation could suggest that a redox shuttle is present in our c-Si cells. In the absence of a shuttle, the reversible self-discharge should correlate with the extent of oxidative side-reactions happening at the positive electrode during storage. For a fixed electrolyte, these reactions are favored at higher cathode potentials. At the end of charge, cathode potentials are larger for Gr cells (~4.29 V vs. $Li/Li^+$) than in c-Si cells at 3.9 V (~3.95 V vs. $Li/Li^+$). Yet, the reversible self-discharge is similar for both systems. Although lithiation of the overhang could also contribute to reversible self-discharge (in case $Li^+$ at the AOH takes longer than one cycle to return to the cathode), this mechanism should also lead to higher self-discharges in Gr than in c-Si, which is not what we observe. We propose that certain species produced at the anode are transported to the cathode where they are re-oxidized, with a net-zero change to the $Li^+$ inventory of the cell. The reversible self-discharge for c-Si cells with varying AOH sizes becomes very similar once normalized by the total anode area, suggesting that the entire surface of the negative electrode would contribute to generate such traveling species. The existence of redox shuttles that are specific to Si-rich cells is an interesting possibility that merits further investigation.

***Extrapolation to other systems of interest.*** The experiments above demonstrated that, in graphite-free c-Si electrodes, the active material at the AOH is not significantly lithiated, be it during slow cycling or extended storage at elevated temperatures. The phase transition that occurs during the very first time c-Si particles lithiate offers the means to test this hypothesis, justifying our initial focus on this type of material. A more general approach to examine overhang utilization is the inspection of coulombic efficiency trends.

Figure 8 shows the CEs at an arbitrary point in the life of several types of full-cells. The data were acquired during testing that involved aging at rates between C/3 and 1C (see Section S4 of the *Supplementary Material*), and occasional slow reference performance tests (RPTs) at $\leq$ C/10 (indicated by red triangles). Whether or not the active material at the overhang is accessible to $Li^+$ can be inferred from the evolution of CEs when the cycling rate is changed. Consider the Gr-based cells in Figures 8a,b, for example. Upon resuming the faster aging cycles after every RPT, the CE is lower and trails over several cycles until returning to its regular progression. This trailing has been linked with utilization of the AOH, and we have observed this behavior in hundreds of similar cells over the years. Long et al. used a physics-based model to analyze this phenomenon and reported that the drop in CE is caused by how access to the AOH varies with cycling rate.[14] They showed that the overhang can be delithiated more effectively during the slower cycles, causing a relative depletion of Li$^+$ after every RPT. When faster cycling is resumed, a fraction of the capacity charged into the anode replenishes the overhang with Li$^+$, causing a temporary decrease in the measured CE until a steady state is once again recovered. An example that supports this analysis is provided in Figure 8c, in which voltage holds were applied at the end of every discharge step of this Gr cell. In this case, the degree of emptying of the overhang remains approximately consistent across all cycles, eliminating distortions to CE values. We believe that this trailing of the CE at alternating cycling rates is a useful method to infer whether or not the active material at the overhang is being lithiated to a meaningful extent, and next we will apply it to Si-containing cells we have tested over the last several years.

Using this method in cells containing graphite-free c-Si electrodes (Figures 8d,e) led to no indication of AOH lithiation, in agreement with the results discussed above. (Note that these cells exhibited growing CE during the initial 100 cycles, explaining the initial trends.) The same was observed for cells using graphite-free SiO$_x$ anodes (Figures 8f,g). The crystallinity of Si domains in SiO$_x$ materials can vary depending on processing history,[32] and the one we used behaves as a-Si (Figure S13). Combining this same SiO$_x$ powder with varying amounts of Gr (< 20wt% Gr) did not change the CE trends (Figures 8h-j). All the evidence above seems to suggest that, just like c-Si, AOH utilization in a-Si electrodes may not occur to a meaningful extent. This is further supported by data in Figure S14, showing experiments analogous to Figure 6 but with an electrode containing a-Si nanoparticles. Among all SiO$_x$ systems we have evaluated in the past, two notable exceptions exist (Figures 8k,l). These cells contained anodes with the same composition as used in Figures 8h,i,k,l and yet they exhibited traits that are indicative of AOH lithiation. Although we have not been able to decisively determine why they differed from other examples, we speculate that it could be related to variations in the effective potentials experienced by the anodes at the end of lithiation in each case. All in all, we leave this as a useful example that, although less likely than in Gr, utilization of the overhang is certainly possible in a-Si electrodes (or similarly-behaved $SiO_x$, as in the present case). Additional information and testing data for cells featured in Figure 8 are provided in Section S4 of the *Supplementary Material*.

Several other variables could be important to determine whether the AOH would be utilized in Si electrodes paired with layered oxide cathodes. Certain Si-rich cells are designed with elevated capacity ratios between the NE and the PE (N/P ratio),[33] constraining Si particles to relatively high electrical potentials and further limiting the driving force for AOH equalization. Prelithiation can also have a significant effect depending on how it is implemented, as it could amorphize c-Si particles over the entire anode and make equalization more likely. Yet another factor is the phase behavior of the lithiated Si particles. Although Si materials are generally expected to form a crystalline $Li_{15}Si_4$ phase when sufficiently lithiated, nucleation of this phase can be kinetically frustrated by the stresses within the electrode.[34] One hallmark of this phase is that, whereas amorphous $Li_xSi_y$ starts delithiating at ~130 mV vs. $Li/Li^+$, lithium extraction from $Li_{15}Si_4$ only commences at > 0.4 V vs. $Li/Li^+$,[34, 35] increasing the voltage hysteresis of the anode and hence the thermodynamic barriers for equalization.

Composite electrodes containing Si-Gr blends are also relevant systems to consider, given the progressive growth in the Si content of anodes used in high-energy commercial cells. Although cells with a high Si content tend to behave as Gr-free systems (as in Figures 8h-j), we hypothesize that the behavior may change at increasing Gr ratios. In this case, the plateaus of Gr could have an interesting role in "pinning" the potential to help drive AOH utilization. Figure 9 shows voltage profiles for cells with 15 wt% Si and 73 wt% Gr. When cells rest after lithiation to 50 mV, the voltage will relax to the OCV of Gr's stage 1 plateau (~88 mV). As $Li^+$ transits from the AA to the AOH, the anode potential will remain nearly unchanged for as long as $LiC_6$ persist in the AA, maintaining the driving force for equalization. Consequently, not only the Gr flakes at the AOH could be accessible to $Li^+$, but also the Si particles. At a high enough Gr content, this pinning effect would also enable lithiation of c-Si at the overhang (Figure S16), since the stage 1 plateau is at a lower potential than the required for the initial formation of $Li_xSi_y$.

Finally, one way to decisively force $Li^+$ into the overhang of c-Si is to cycle it versus a counter electrode presenting a flat voltage profile and sufficient $Li^+$ inventory (such as Li metal, and LFP in certain cells). Differently from layered oxide cathodes, these electrodes can provide large amounts of lithium ions at a given cell voltage (as electrode delithiation does not cause an increase in electrode potential). Cells containing a 11 mm Li metal vs. 14 mm c-Si (38% AOH) showed full utilization of the overhang after C/10 cycling with a voltage hold at 50 mV until the current decreased to C/20 (< 1 hour, Figure S17). Interestingly, the overhang appeared to be readily accessible during the constant current cycling, since much more capacity was accessible in the electrode than expected for the AA region alone (Figure S18). We expect that a similar behavior could occur in LFP cells, especially in experiments involving extended voltage holds.[19]

*Conclusions*

Commercial Li-ion batteries contain excess area of the negative electrode. Although this overhang will experience limited direct $Li^+$ flux during charging, capacity exchanges over time with the active area of graphite electrodes causes these appendages to be highly active. Given the relatively long distances that $Li^+$ must travel (of the order of centimeters), these exchanges are quite slow and can extend over several months. Hence, much of the testing data collected for Li-ion batteries may be distorted by the behavior of the overhang, which can lead to either over- or underestimation of aging rates, depending on cell history. Most life forecasting models do not explicitly account for these effects, and hence they may only be applicable to systems in which the overhang behaves like in traditional graphite electrodes.

In the present work we showed that, in many cases, Si is not such a system. Overhang equalization during cell rest relies on the concerted delithiation from the AA and lithiation of the AOH without external energy input. The large voltage hysteresis of silicon imposes a large energy penalty for delithiation, decreasing the reducing power that can be achieved when lithiating the overhang. If Si particles are initially crystalline, this barrier becomes even larger due to the necessity of reaching ~110 mV for lithiation to initiate, effectively imposing a thermodynamic barrier for access to the overhang. Using Raman spectroscopy, we demonstrate this to be the case even when full-cells are stored at 45 $^{\circ}$C for 4 months after the anode was lithiated to ~50 mV vs. Li/Li$^+$. Curiously, we observed that, despite the inactivity of c-Si particles at the AOH, the capacity fade of cells during calendar aging was higher for increasing overhang sizes. We demonstrate that the carbon particles that are part of the conductive network of the electrode are electrochemically active, being able to both store charge and carry electrons to the overhang, leading to SEI growth outside the bounds of the active area.

We also suggested that analysis of coulombic efficiency trends during cycling can be used as indicator of overhang utilization, as changes in the Li$^+$ content of the AOH when switching rates will cause the CE to trail over several cycles. We used this method to analyze data from cells containing several types of Si electrodes and found that overhang equalization in amorphous Si anodes appears to be unlikely, but possible under some scenarios. Furthermore, for Si-Gr electrodes with a low Si content, we showed that AOH equalization appears to proceed unimpeded.

The use of Si-rich electrodes can increase cell energy but may lead to changes in physical processes occurring in the battery, such as the case of overhang equalization. Though these changes may appear to be subtle at first glance, they can significantly impact attempts to quantify present and future states of aging in these cells.

*Acknowledgements*

The authors are grateful to Kevin Knehr and Ankit Verma for valuable discussions. This research was supported by the U.S. Department of Energy's Vehicle Technologies Office under the Silicon Consortium Project, directed by Brian Cunningham, Thomas Do, Nicolas Eidson and Carine Steinway, and managed by Anthony Burrell. This work was supported in part by the U.S. Department of Energy, Office of Science, Office of Workforce Development for Teachers and Scientists (WDTS) under the Science Undergraduate Laboratory Internships Program (SULI). A portion of this work (silicon milling – GMV) was performed at ORNL operated by UT-Battelle, LLC, under contract DEAC05-00OR22725 with the U.S. Department of Energy (DOE). The submitted manuscript has been created by UChicago Argonne, LLC, Operator of Argonne National Laboratory ("Argonne"). Argonne, a U.S. Department of Energy Office of Science laboratory, is operated under Contract No. DE-AC02-06CH11357. The U.S. Government retains for itself, and others acting on its behalf, a paid-up nonexclusive, irrevocable worldwide license in said article to reproduce, prepare derivative works, distribute copies to the public, and perform publicly and display publicly, by or on behalf of the Government.

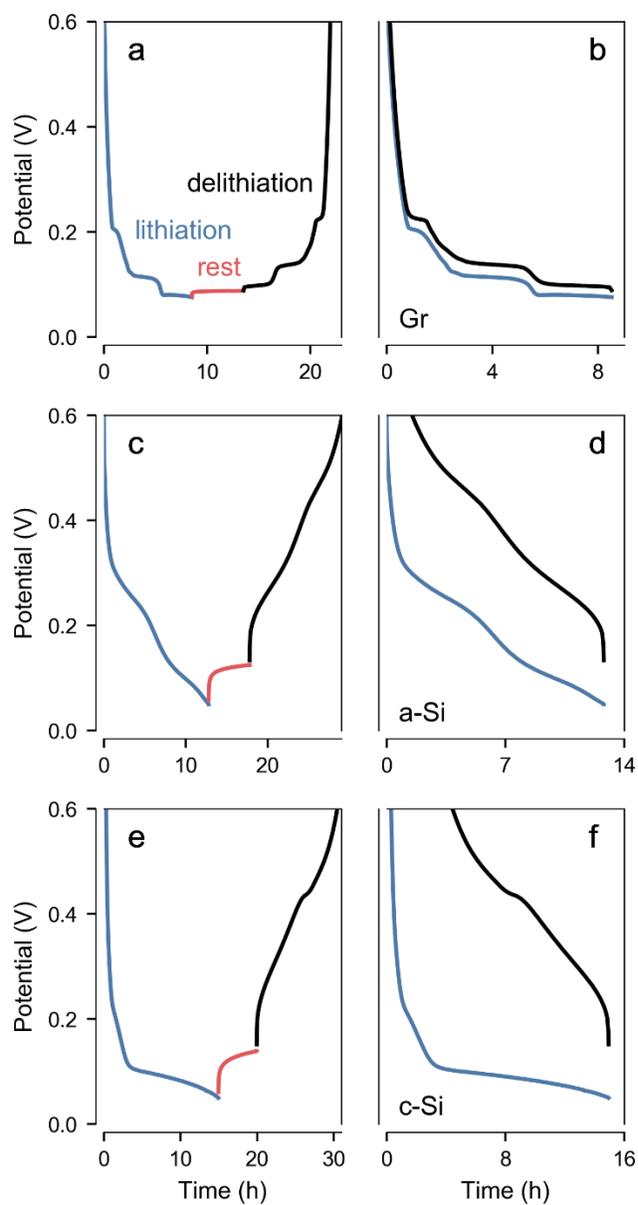

Figure 1. The hysteresis of various types of negative electrodes measured at C/10 in cells vs. Li metal: a,b) graphite; c,d) graphite-free amorphous Si; e,f) graphite-free crystalline Si. This hysteresis decreases the driving force for overhang equalization. Data for a-Si came from a later cycle of the c-Si cell. The color code defined in panel *a* also applies to panels *c* and *e*.

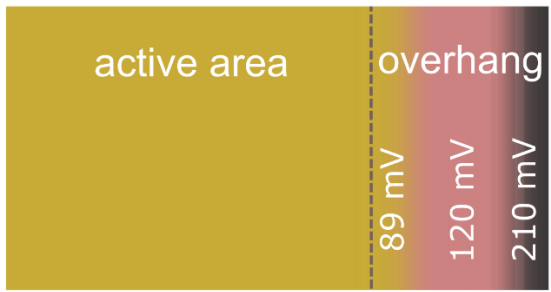

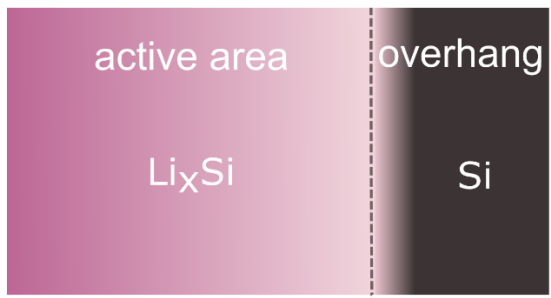

Figure 2. Schematic representation of Li$^+$ propagation into the overhang of graphite (*top*) and silicon (*bottom*) electrodes. The low hysteresis and the plateaus in the voltage profile of graphite establish a driving force for homogenizing the electrical potentials between the active area and the overhang, prompting Li$^+$ exchange to take place. Overhang regions closer to the active area are generally observed to lithiate faster. In the case of silicon, Li$^+$ transfer to the overhang is inhibited by the large voltage hysteresis, which acts as a thermodynamic barrier that decreases the driving force for equalization. When lithiation of the overhang is possible, a decrease in the Li$^+$ content of the surrounding portions of the active area will lead to even lower driving forces, causing the equalization process to be highly inefficient.

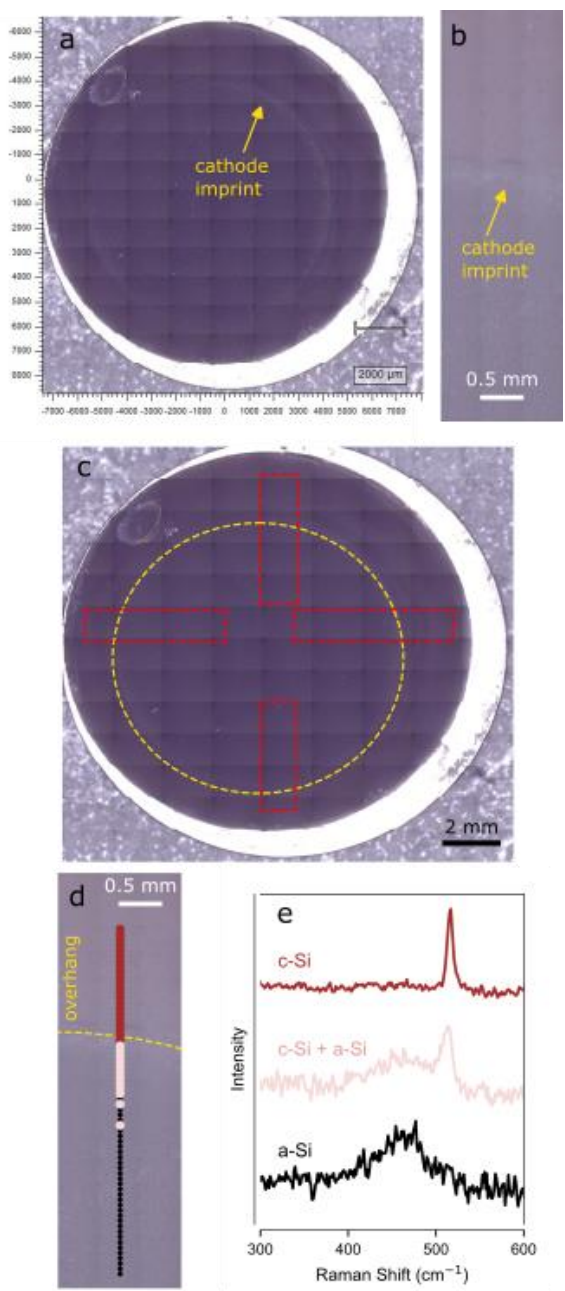

Figure 3. Overview of the methodology for presenting Raman data: a) optical image obtained with a 5x objective of an entire cycled anode from a cell with 96% overhang. A faint circular imprint left by compression against the cathode is visible; b) optical image obtained with a 20x objective of a smaller region of the electrode. The position of the overhang, as established by the cathode imprint, is still visible. c) high-resolution version of image in panel *a*, with the position of the cathode marked in yellow and the areas probed in detail marked in red; d) example of a Raman map. Circles mark the points were spectra were taken, with colors indicating the outcomes (color code in panel *e*). The cathode imprint is highlighted in yellow; e) example of Raman spectra and the color code used for data presentation. *Black* indicates that only amorphous Si is present, *red* that only crystalline Si is detected and *pink* that bands from both are observed.

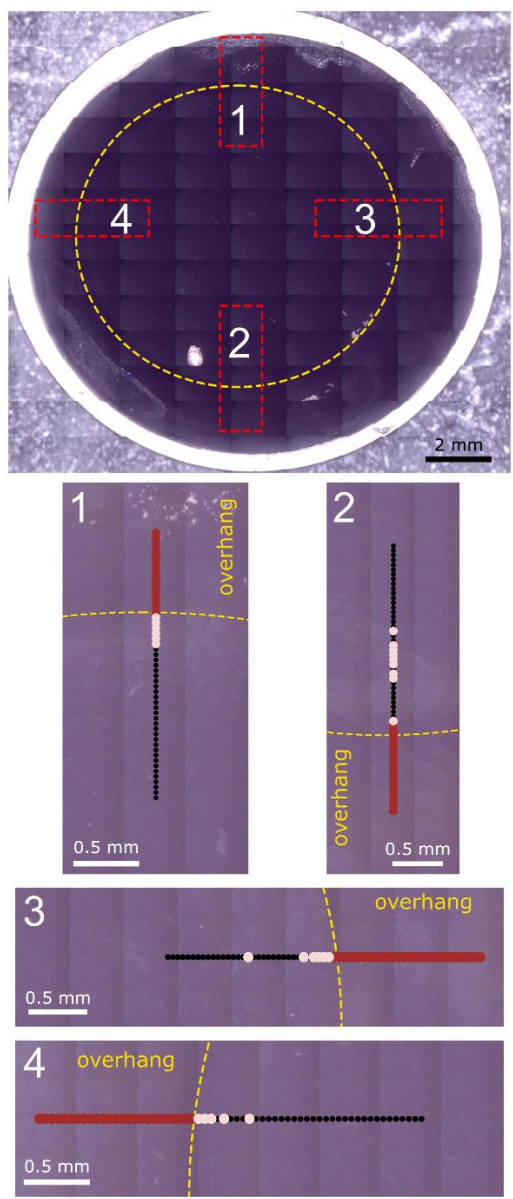

Figure 4. Raman data from the c-Si anode of a full-cell with 96% overhang that was charged to 3.9 V (~50 mV vs. Li/Li$^+$ at the negative electrode) and stored at 45 $^o$C for four months. The numbers in the top image are used to indicate where the corresponding Raman maps were taken. Only crystalline Si is found at the overhang, indicating that these regions were never lithiated.

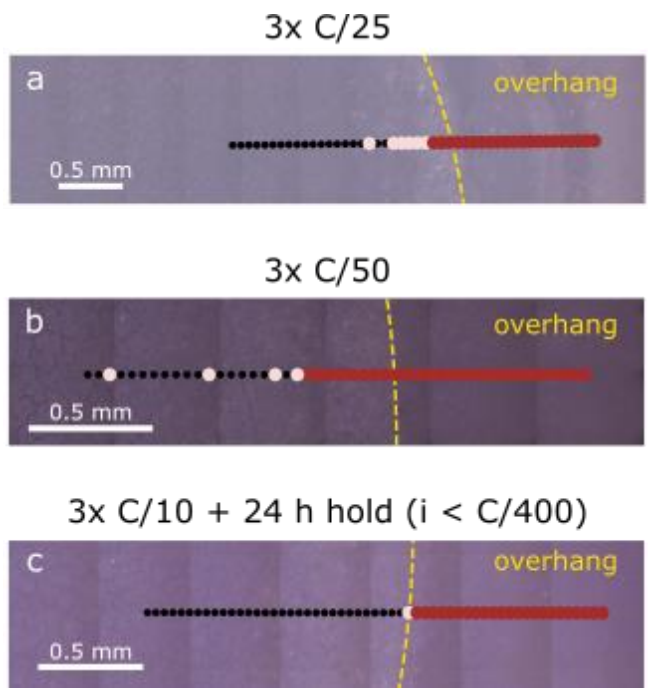

Figure 5. Example of Raman maps obtained from c-Si anodes of full-cells with 96% overhang exposed to different cycling protocols: a) three full cycles at C/25; b) three full cycles at C/50; c) three full cycles at C/10, with a 24-hour voltage hold at the top of charge. The upper cutoff voltage was 3.9 V (~50 mV vs. Li/Li$^+$ at the negative electrode). Slower rates can increase the utilization of the active area but appears to be insufficient to prompt lithiation of the overhang.

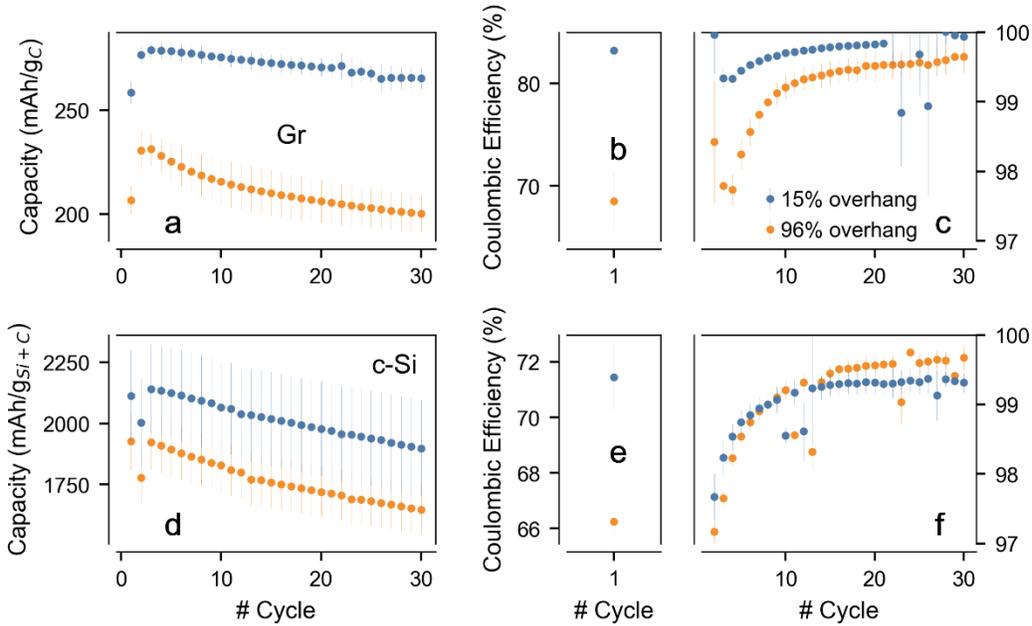

Figure 6. Effect of overhang size on cell behavior during cycling at C/10: a) capacity; and b,c) coulombic efficiency for graphite vs. NMC811 cells; d) capacity; and e,f) coulombic efficiency for c-Si vs. NMC811 cells. Blue and orange markers denote cells with 15% and 96% overhang, respectively. Unplanned test interruptions affected the progression of coulombic efficiencies (blue markers after cycle 20 in panel *c*, and both types of cells in panel *f*). Capacities are normalized by the weight of materials within the active area only. The legend in panel *c* applies to all panels.

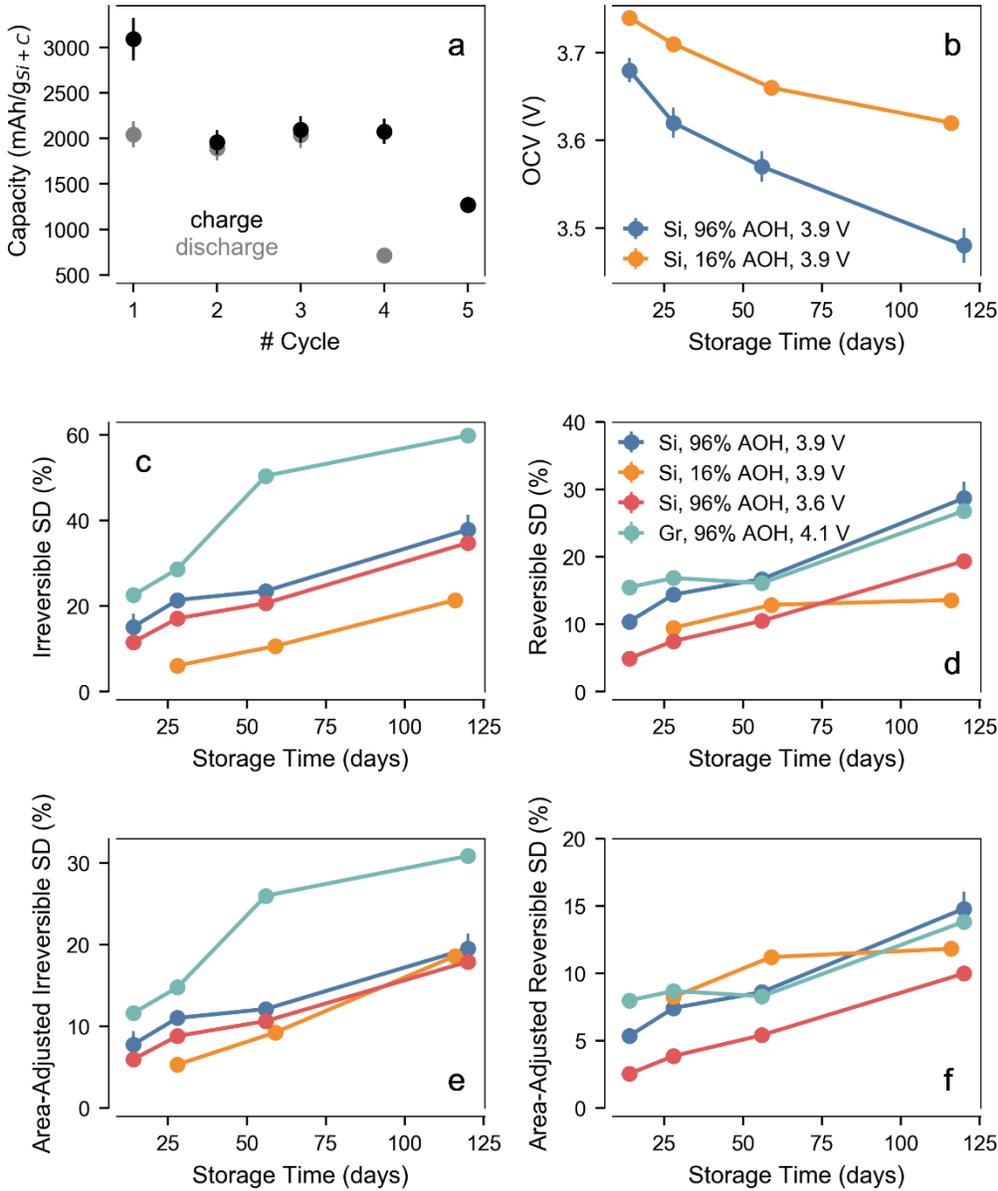

Figure 7. Electrochemical observations from calendar aging experiments: a) Example of capacity data from an aging experiment. Cells are stored at 45 °C after the fourth charge. b) Voltage drop measured after storing c-Si vs. NMC811 cells at 45 °C for varying periods. c) irreversible self-discharge; and d) reversible self-discharge observed after aging; e) like panel *c* but normalized by the area of the anode; f) like panel *d* but normalized by the area of the anode. Legend in panel *d* also applies to panels *c,e,f*.

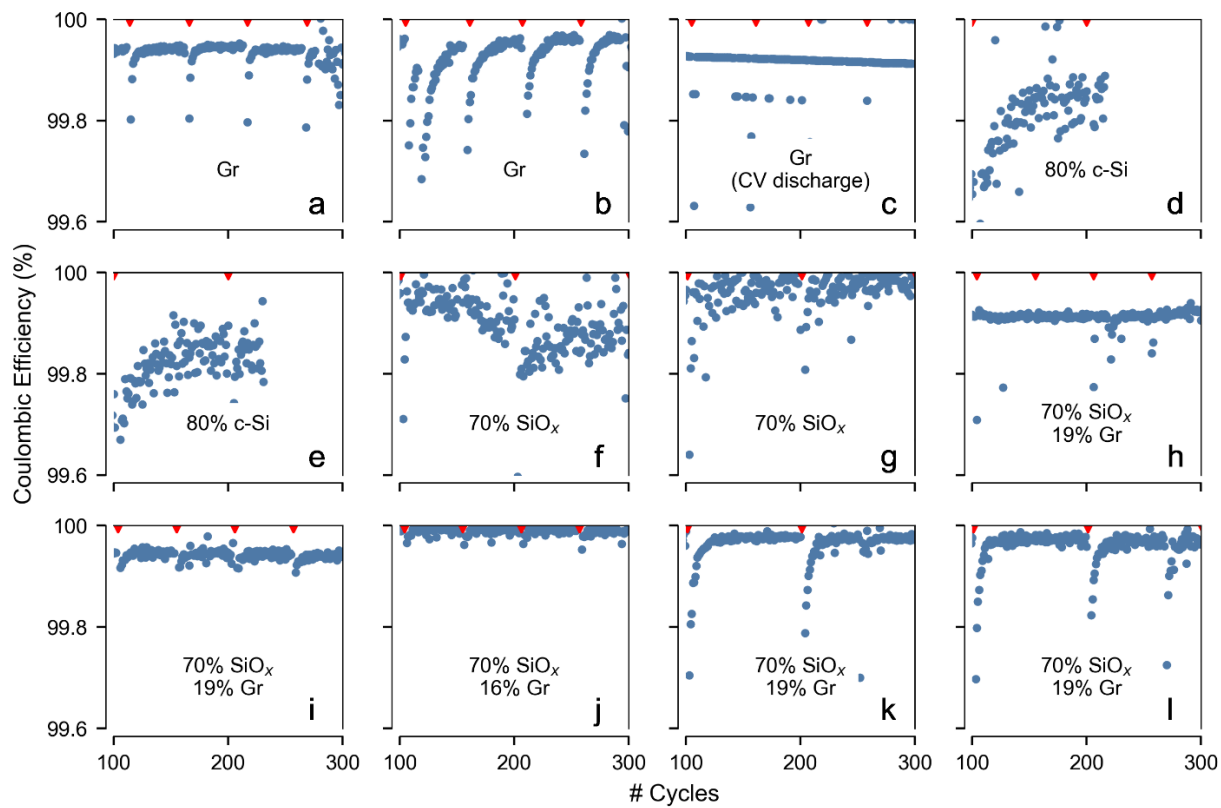

Figure 8. Coulombic efficiencies at an arbitrary point in the life of full-cells cells using the indicated anodes. Slow reference performance tests were taken at the points indicated by the red triangles shown at the top of each panel. Occasional unplanned test interruptions (and failure of the temperature chamber in the later cycles of panel *a*) increased the dispersion in some of the data points. Refer to the *Supplementary Material* for complete information about the cells.

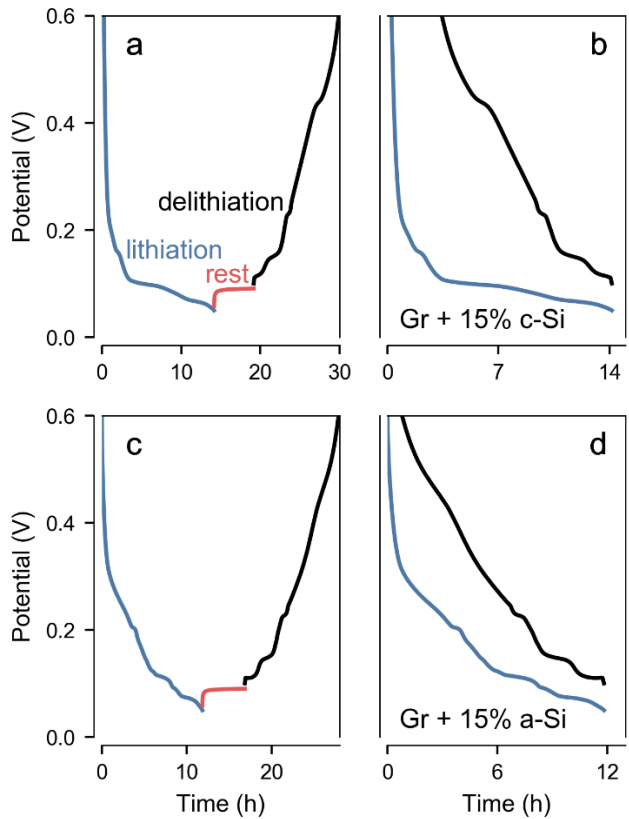

Figure 9. Hysteresis in Si-graphite electrodes (15wt% Si) measured vs. Li metal: a,b) crystalline Si; c,d) amorphous Si. The graphite plateaus can pin the anode potential and help drive overhang equalization. Data for a-Si came from a later cycle of the c-Si cell. The color code defined in panel *a* also applies to panel *c*.

*Supplementary Material*

# Inactive Overhang in Silicon Anodes

Aidin I. O'Brien,[1] Stephen E. Trask,[1] Devashish Salpekar,[1] Seoung-Bum Son,[1] Alison R. Dunlop,[1] Gabriel M. Veith,[2] Wenquan Lu,[1] Brian J. Ingram,[1] Daniel P. Abraham,[1] Andrew N. Jansen,[1] Marco-Tulio F. Rodrigues[1*]

[1] Chemical Sciences and Engineering Division, Argonne National Laboratory, Lemont, IL, USA

[2] Chemical Sciences Division, Oak Ridge National Laboratory, Oak Ridge, TN, USA

**\*Contact:** marco@anl.gov

## Section S1. Supporting figures from calendar aging studies

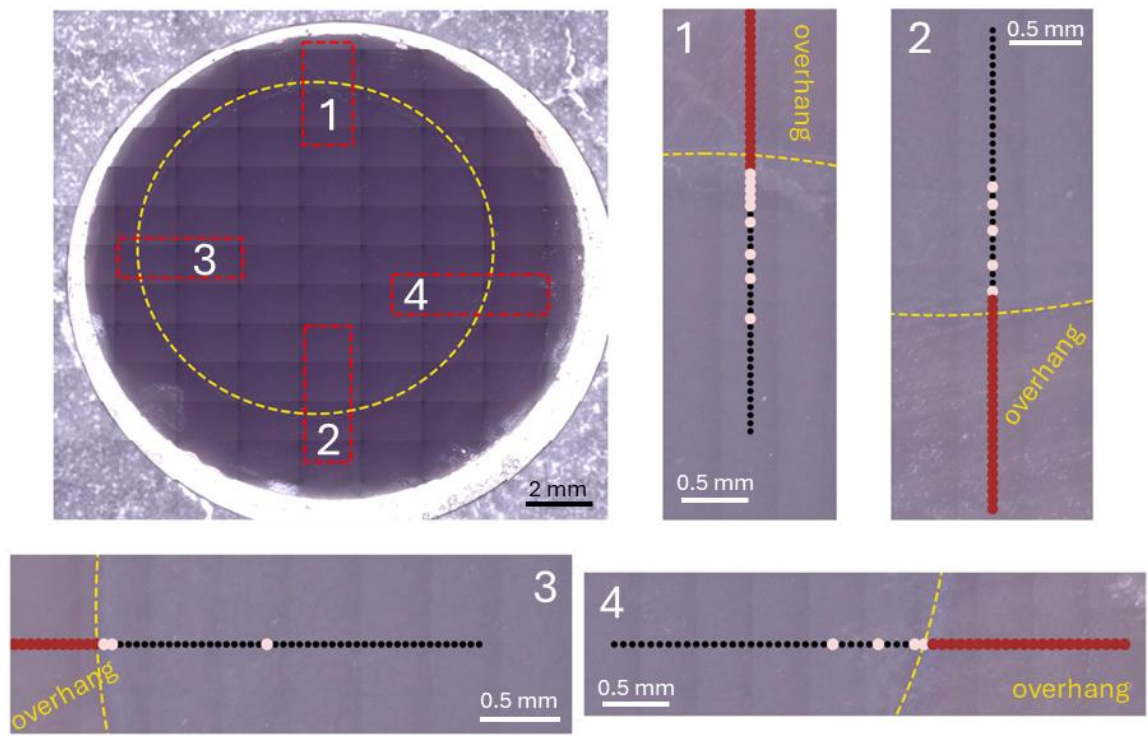

Figure S1. Raman studies with a c-Si electrode extracted from a full-cell with 96% overhang that had been charged to 3.9 V (~50 mV vs. Li/Li$^+$ at the anode) and stored at 45 $^o$C for 1 month.

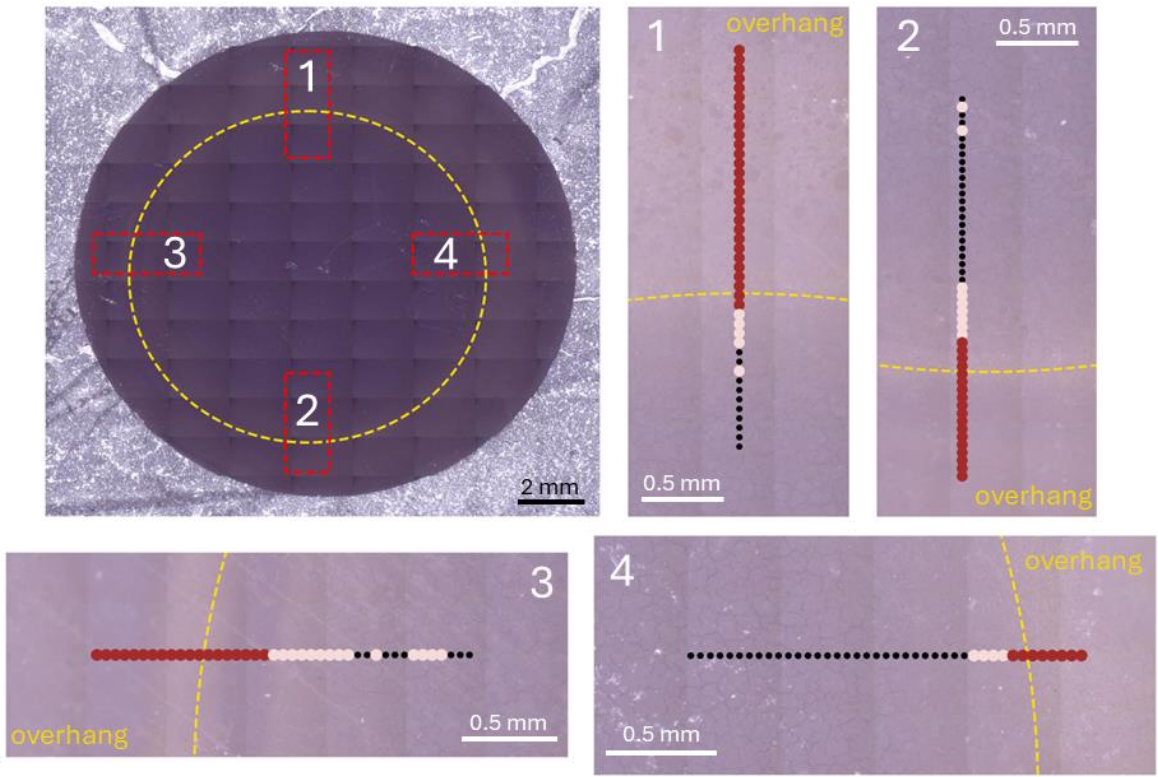

Figure S2. Raman studies with a c-Si electrode extracted from a full-cell with 96% overhang that had been charged to 3.9 V (~50 mV vs. Li/Li$^+$ at the anode) and stored at 45 $^o$C for 2 months.

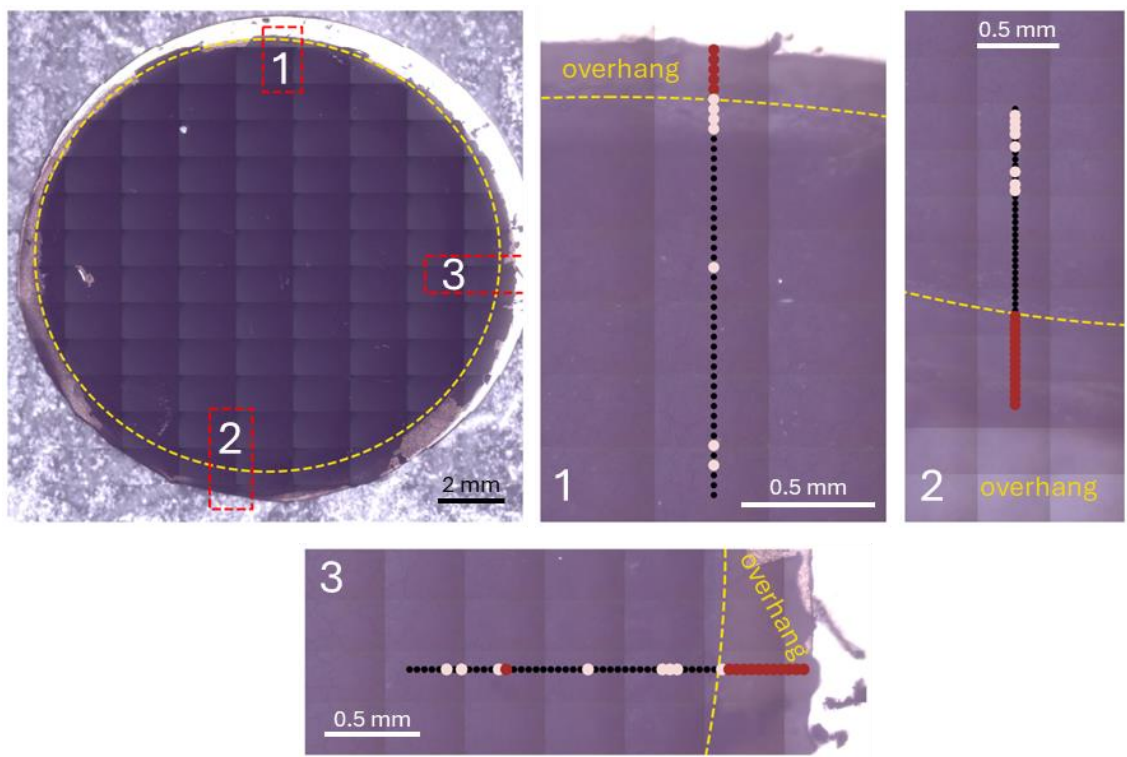

Figure S3. Raman studies with a c-Si electrode extracted from a full-cell with 15% overhang that had been charged to 3.95 V (<50 mV vs. Li/Li$^+$ at the anode) and stored at 45 °C for 2 months. Even these higher levels of lithiation did not lead to significant overhang utilization. The electrode experienced slight delamination and deformation when handled during disassembly.

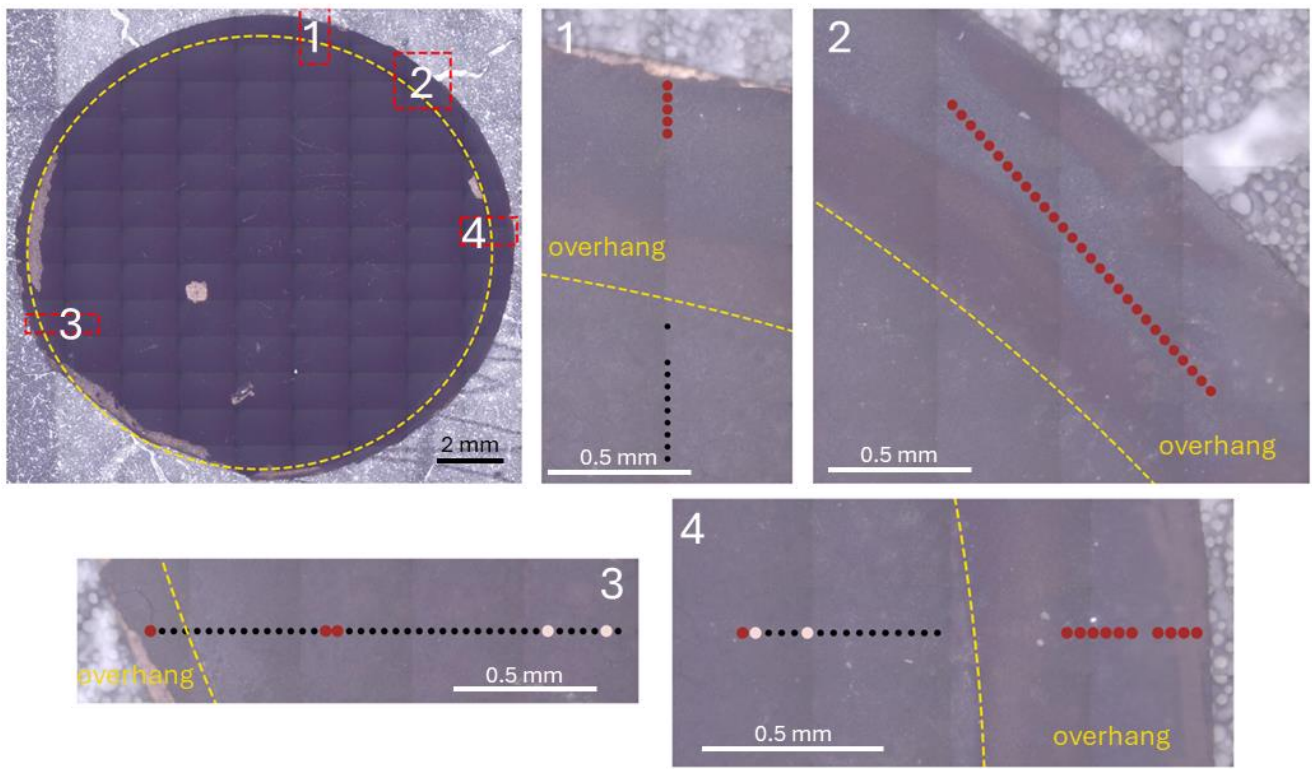

Figure S4. Raman studies with a c-Si electrode extracted from a full-cell with 15% overhang that had been charged to 3.9 V (~50 mV vs. Li/Li$^+$ at the anode) and stored at 45 $^{\circ}$C for 4 months. Electrode deformation during disassembly made it difficult to retain focus over extended areas. We opted for showing this data but retaining only the data points where Raman features can be more clearly discerned.

## Section S2. Supporting figures from formation studies

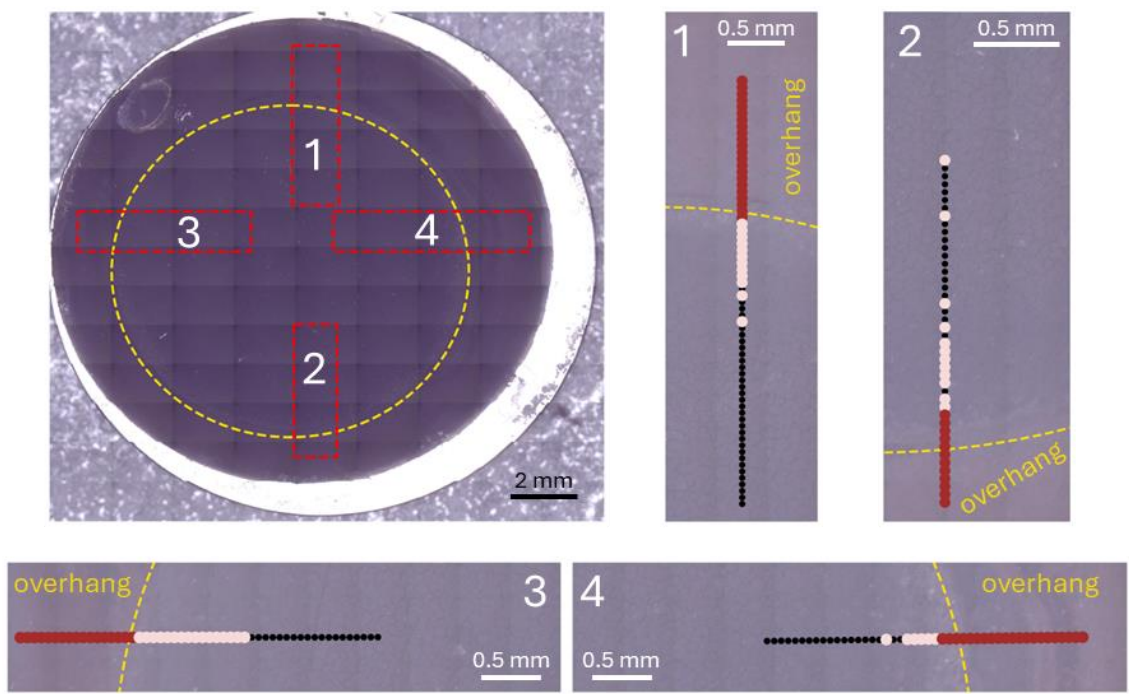

Figure S5. Raman studies with a c-Si electrode extracted from a full-cell with 96% overhang after 3x cycles at C/25. This is the complete study for data shown in Figure 5a.

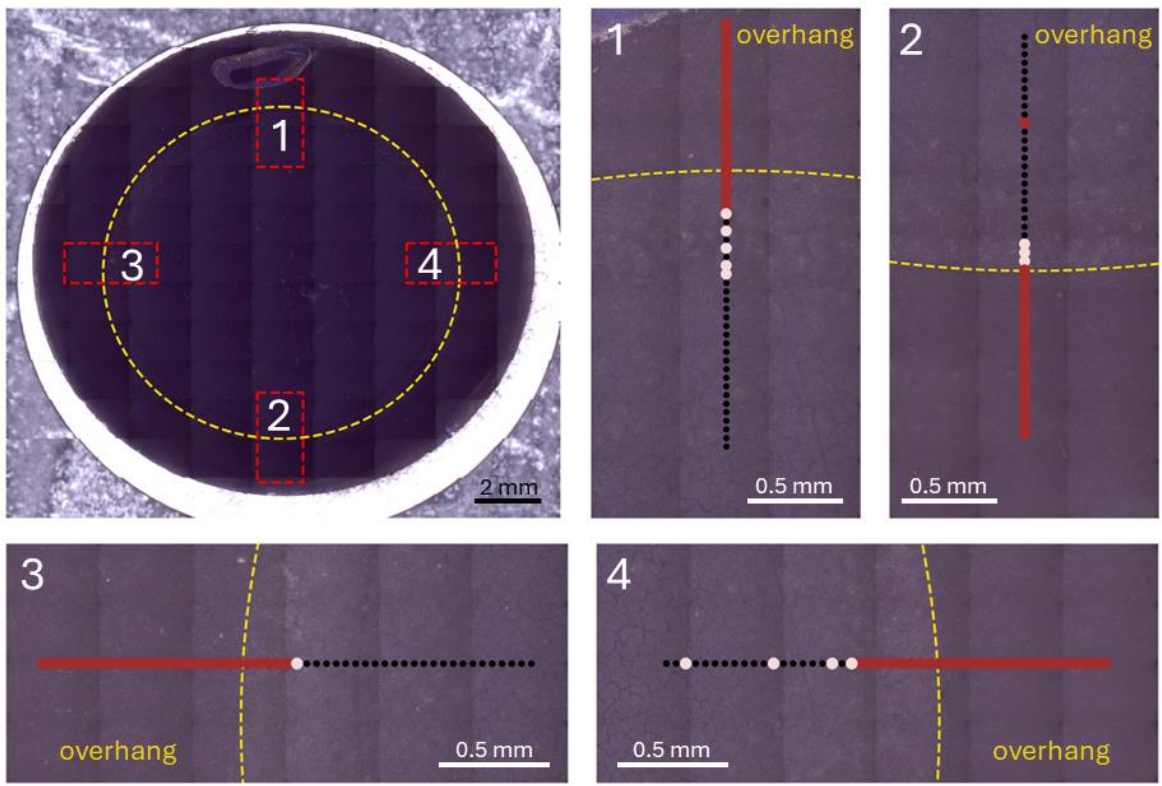

Figure S6. Raman studies with a c-Si electrode extracted from a full-cell with 96% overhang after 3x cycles at C/50. This is the complete study for data shown in Figure 5b.

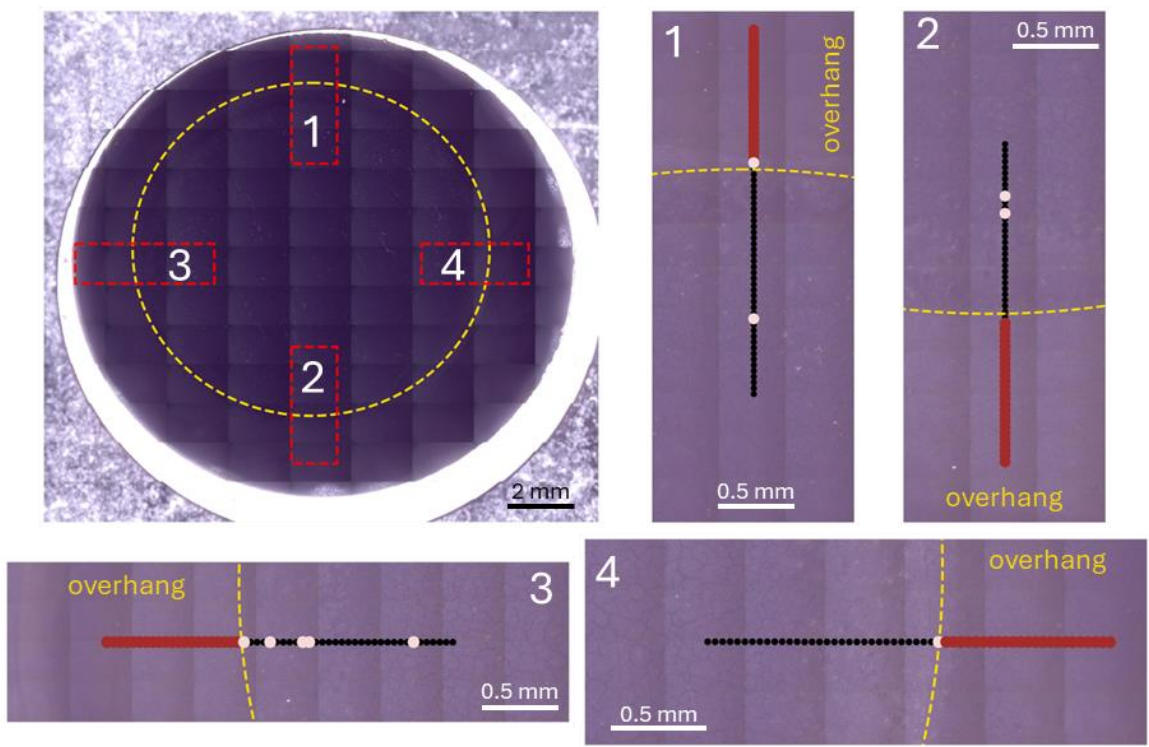

Figure S7. Raman studies with a c-Si electrode extracted from a full-cell with 96% overhang after 3x cycles at C/10 with 24-h-long voltage holds at the top of charge. This is the complete study for data shown in Figure 5c.

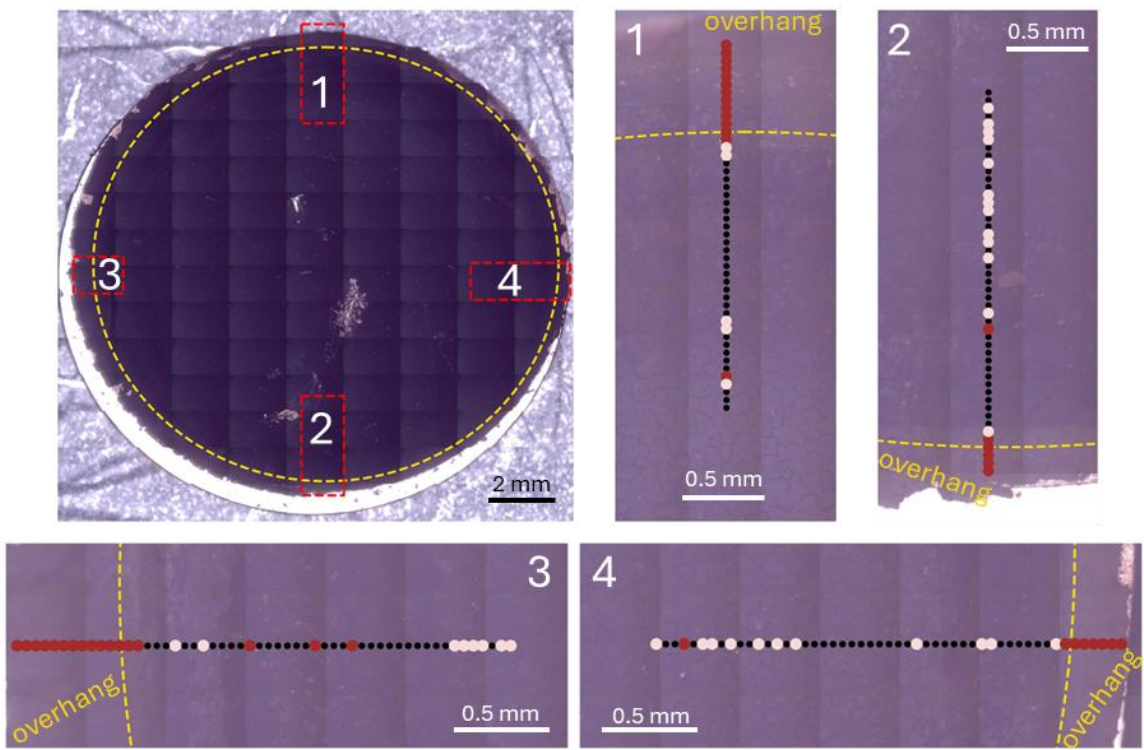

Figure S8. Raman studies with a c-Si electrode extracted from a full-cell with 15% overhang after 3x cycles at C/10 with 24-h-long voltage holds at the top of charge. The electrode experienced slight delamination when handled during disassembly.

## Section S3. Supporting figures from slow cycling studies

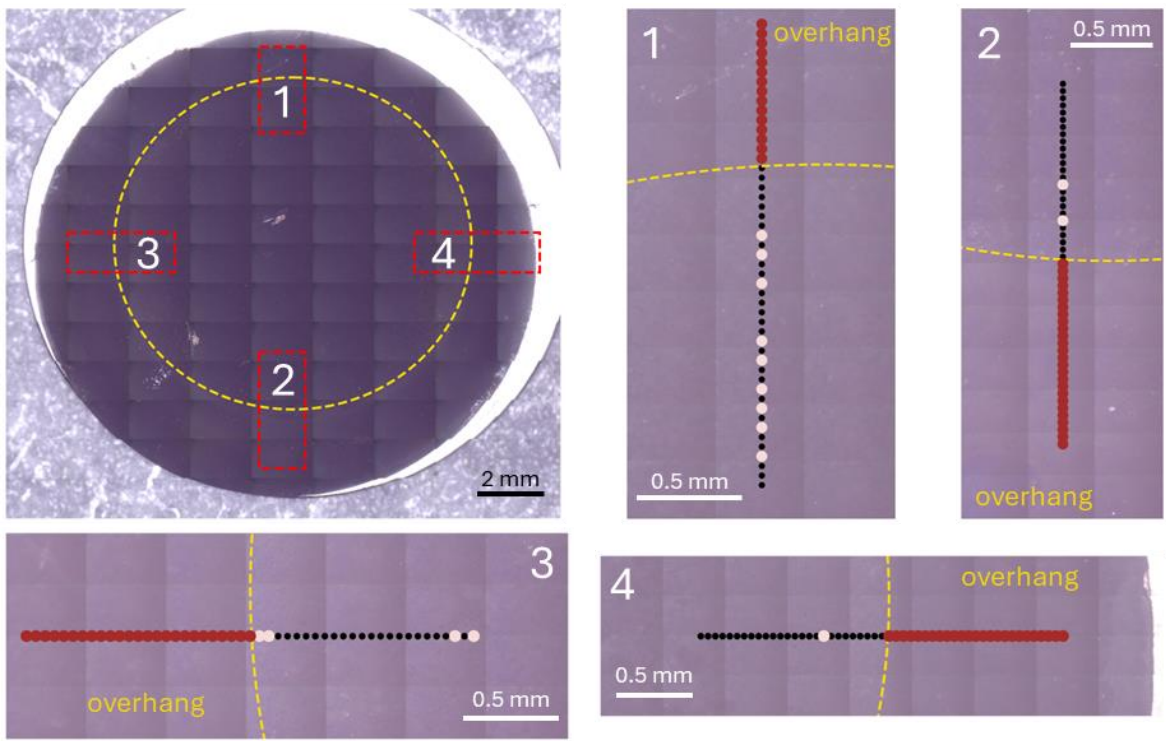

Figure S9. Raman studies with a c-Si electrode extracted from a full-cell with 96% overhang after 30x cycles at a slow C/10 rate. The electrode is from one of the cells featured in Figure 6d.

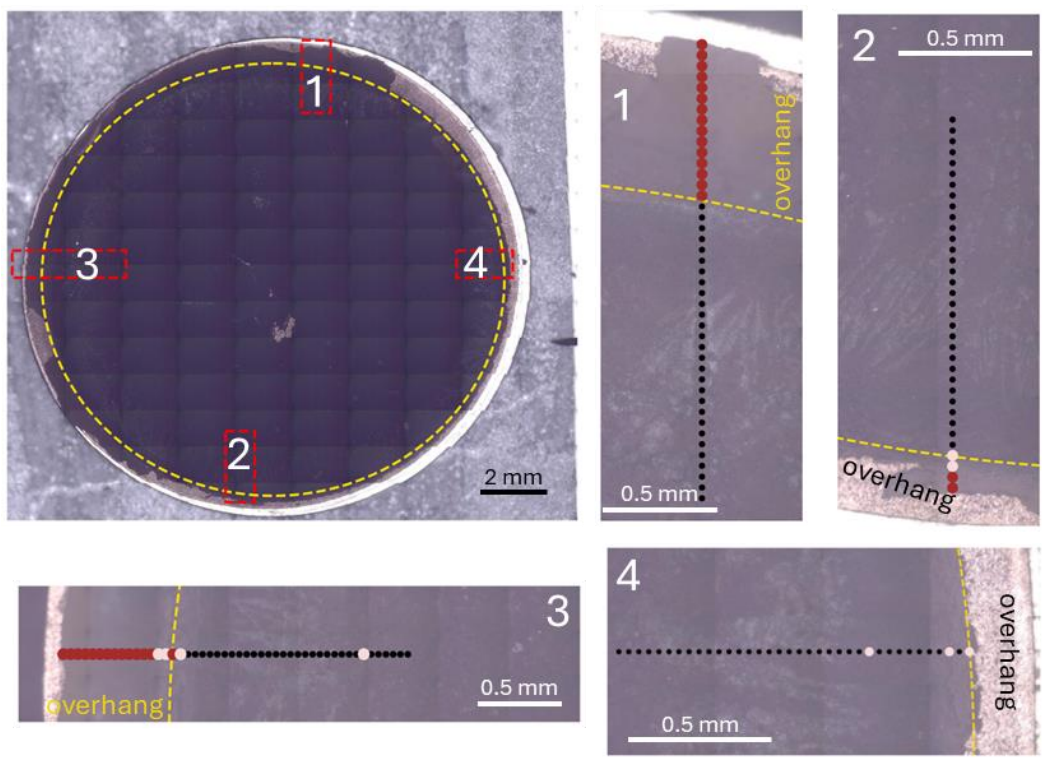

Figure S10. Raman studies with a c-Si electrode extracted from a full-cell with 15% overhang after 30x cycles at a slow C/10 rate. The electrode is from one of the cells featured in Figure 6d. The electrode experienced slight delamination when handled during disassembly.

# Section S4. Additional information for tests shown in Figure 8

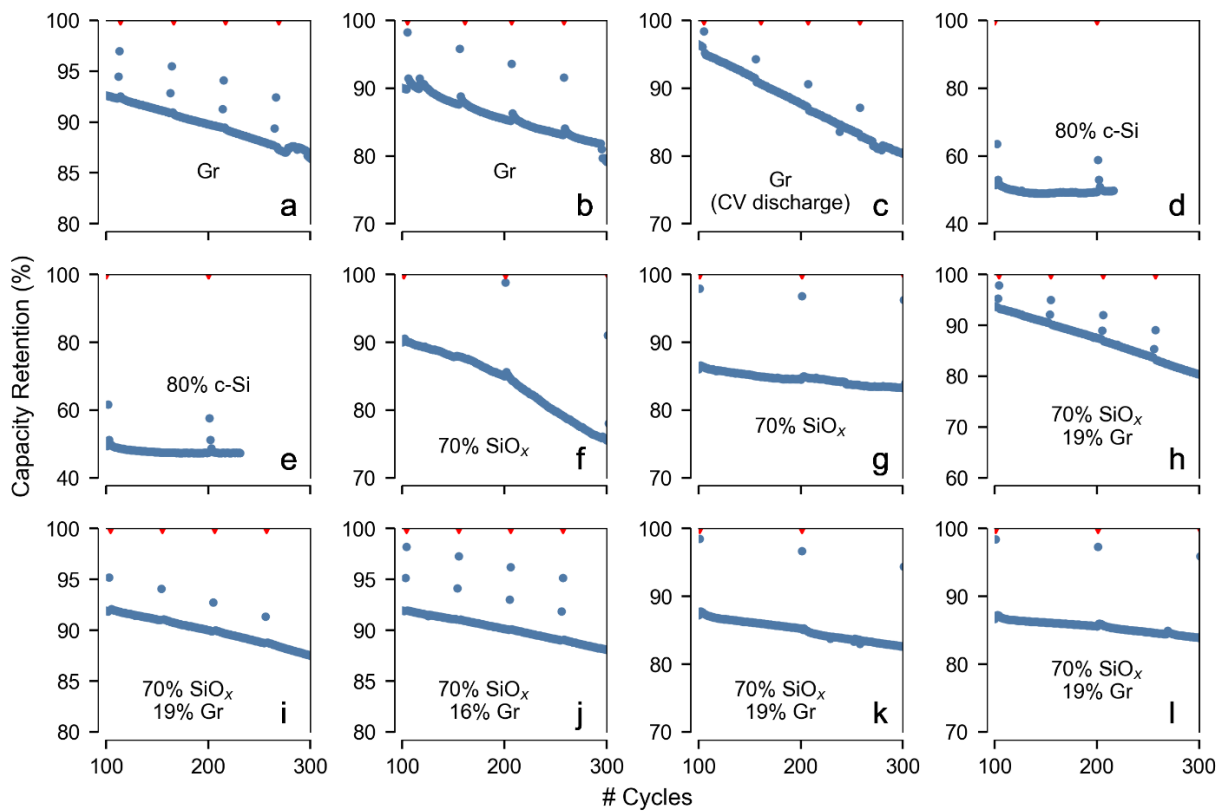

Figure S11. Corresponding capacities for cells shown in Figure 8. The position of RPTs is indicated by the red triangles at the top of the image.

Table S1. Electrode composition for cells for which data are presented in Figures 8 and S11.

| **NE1: Gr** <br> 91.83wt% Superior Graphite SLC1506T, 2wt% Timcal C45 carbon, 6wt% Kureha 9300 PVDF binder, 0.17wt% oxalic acid; 47 µm coating thickness; 37.4% porosity; 6.38 mg/cm$^2$ coating | **PE1: NMC955** <br> 90wt% NMC 90-5-5, 5wt% Timcal C45, 5wt% Solvay 5130 PVDF; 34 µm coating thickness; 32.7% porosity; 9.40 mg/cm$^2$ coating |
|---|---|
| **NE2: Gr** <br> 91.83wt% Superior Graphite SLC1506T, 2wt% Timcal C45 carbon, 6wt% Kureha 9300 PVDF binder, 0.17wt% oxalic acid; 70 µm coating thickness; 38.2% porosity; 9.38 mg/cm$^2$ coating | **PE2: NMC532** <br> 90wt% Toda NMC532, 5wt% Timcal C45, 5wt% Solvay 5130 PVDF; 80 µm coating thickness; 42.8% porosity; 18.57 mg/cm$^2$ coating |
| **NE3: c-Si** <br> 80wt% milled c-Si, 10wt% Timcal C45 carbon, 10wt% polyimide binder; 20 µm coating thickness; 52% porosity; 1.60 mg/cm$^2$ coating | **PE3: NMC811** <br> 96wt% Targray NMC811, 2wt% Timcal C45, 2wt% Solvay 5130 PVDF; 72 µm coating thickness; 34.7% porosity; 21.01 mg/cm$^2$ coating |
| **NE4: SiO$_x$** <br> 70wt% Osaka SiO$_x$, 10wt% Timcal C45 carbon, 10wt% LiPAA (Sigma-Aldrich, titrated); 25 µm coating thickness; 49.8% porosity; 2.51 mg/cm$^2$ coating | **PE4: NMC622** <br> 90wt% Ecopro NMC622, 5wt% Timcal C45, 5wt% Solvay 5130 PVDF; 49 µm coating thickness; 33.5% porosity; 13.34 mg/cm$^2$ coating |
| **NE5: SiO$_x$** <br> 70wt% Osaka SiO$_x$, 10wt% Timcal C45 carbon, 10wt% polyimide binder; 19 µm coating thickness; 53.1% porosity; 1.72 mg/cm$^2$ coating | **PE5: NMC622** <br> 90wt% Ecopro NMC622, 5wt% Timcal C45, 5wt% Solvay 5130 PVDF; 35 µm coating thickness; 31.7% porosity; 9.78 mg/cm$^2$ coating |
| **NE6: SiO$_x$-Gr** <br> 69.98wt% Osaka SiO$_x$, 19.07wt% MagE3 graphite, 0.99wt% Tuball SWCNTs, 0.99wt% CMC, 8.98wt% acrylate binder BA-290S (Blue Ocean and Blackstone); 36 µm coating thickness; 61.9% porosity; 3.01 mg/cm$^2$ coating | **PE6: NMC811** <br> 96wt% Targray NMC811, 2wt% Timcal C45, 2wt% Solvay 5130 PVDF; 80 µm coating thickness; 34.9% porosity; 23.29 mg/cm$^2$ coating |
| **NE7: SiO$_x$-Gr** <br> 69.98wt% Osaka SiO$_x$, 19.07wt% MagE3 graphite, 0.99wt% Tuball SWCNTs, 0.99wt% CMC, 8.98wt% acrylate binder BA-290S (Blue Ocean and Blackstone); 66 µm coating thickness; 57.2% porosity; 6.19 mg/cm$^2$ coating | **PE7: NMC811** <br> 96wt% Targray NMC811, 1.95wt% Timcal C45, 0.05wt% Tuball SWCNT, 2wt% Solvay 5130 PVDF; 102 µm coating thickness; 34.2% porosity; 29.98 mg/cm$^2$ coating |
| **NE8: SiO$_x$-Gr** <br> 70wt% Osaka SiO$_x$, 16wt% MagE3 graphite, 1wt% Tuball SWCNTs, 1wt% CMC, 12wt% acrylate binder BA-290S (Blue Ocean and Blackstone); 68 µm coating thickness; 57.1% porosity; 6.25 mg/cm$^2$ coating; double-sided (info per side) | **PE8: NMC811** <br> 96wt% Targray NMC811, 1.95wt% Timcal C45, 0.05wt% Tuball SWCNT, 2wt% Solvay 5130 PVDF; 95 µm coating thickness; 34.4% porosity; 27.85 mg/cm$^2$ coating; double-sided (info per side) |

Table S2. Information for cells for which data are presented in Figures 8 and S11. Bullet points indicate, respectively: composition of PE and NE (see Table S1); cell format; electrolyte composition; relative size of the overhang; voltage limits; rates of reference performance tests and of aging cycles; state of prelithiation. When present, prelithiation involved a number of full cycles (either vs. Li metal or vs. NMC811), after which the NE was harvested and used in a fresh full-cell. Indicated voltages are the approximate values (vs. Li metal) to which the prelithiated NE was delithiated prior to harvesting.

| a) NMC 90-5-5 vs. Gr | b) NMC532 vs. Gr | c) NMC532 vs. Gr | d) NMC811 vs. c-Si |
|---|---|---|---|
| ▪ PE1 vs. NE1 | ▪ PE2 vs. NE2 | ▪ PE2 vs. NE2 | ▪ PE3 vs. NE3 |
| ▪ coin cell (2032) | ▪ coin cell (2032) | ▪ coin cell (2032) | ▪ coin cell (2032) |
| ▪ Gen2* | ▪ Gen2 | ▪ Gen2 | ▪ Gen2 + 3wt% FEC |
| ▪ 15% AOH | ▪ 15% AOH | ▪ 15% AOH | ▪ 15% AOH |
| ▪ 3.0 - 4.2 V | ▪ 3.0 - 4.1 V | ▪ 3.0 - 4.1 V | ▪ 3.4 - 4.25 V |
| ▪ 0.04C / 0.33C | ▪ 0.04C / 0.33C | ▪ 0.04C / 0.33C (+ hold) | ▪ 0.1C / 0.33C |
| ▪ not prelithiated | ▪ not prelithiated | ▪ not prelithiated | ▪ not prelithiated |
| **e) NMC811 vs. c-Si** | **f) NMC622 vs. SiO$_x$** | **g) NMC622 vs. SiO$_x$** | **h) NMC811 vs. SiO$_x$-Gr** |
| ▪ PE3 vs. NE3 | ▪ PE4 vs. NE4 | ▪ PE5 vs. NE5 | ▪ PE6 vs. NE6 |
| ▪ coin cell (2032) | ▪ single-layer pouch cell (14.1 cm$^2$) | ▪ single-layer pouch cell (14.1 cm$^2$) | ▪ coin cell (2032) |
| ▪ Gen2 + 3wt% FEC | ▪ Gen2 + 4wt% FEC | ▪ Gen2 + 4wt% FEC | ▪ Gen2 + 3wt% FEC |
| ▪ 15% AOH | ▪ 5.7% AOH | ▪ 5.7% AOH | ▪ 15% AOH |
| ▪ 3.4 - 4.25 V | ▪ 3.0 - 4.15 V | ▪ 3.0 - 4.15 V | ▪ 3.02 - 4.18 V |
| ▪ 0.1C / 0.33C | ▪ 0.1C / 1C | ▪ 0.1C / 1C | ▪ 0.1C / 0.33C |
| ▪ not prelithiated | ▪ ~0.85 V after prelit. | ▪ ~0.85 V after prelit. | ▪ not prelithiated |
| **i) NMC811 vs. SiO$_x$-Gr** | **j) NMC811 vs. SiO$_x$-Gr** | **k) NMC811 vs. SiO$_x$-Gr** | **l) NMC811 vs. SiO$_x$-Gr** |
| ▪ PE7 vs. NE7 | ▪ PE8 vs. NE8 | ▪ PE7 vs. NE7 | ▪ PE7 vs. NE7 |
| ▪ single-layer pouch cell (14.1 cm$^2$) | ▪ multi-layer pouch cell (46.3 cm$^2$) | ▪ coin cell (2032) | ▪ coin cell (2032) |
| ▪ Gen2 + 3wt% FEC | ▪ Gen2 + 3wt% FEC | ▪ Gen2 + 3wt% FEC | ▪ Gen2 + 3wt% FEC |
| ▪ 5.7% AOH | ▪ 6% AOH | ▪ 15% AOH | ▪ 15% AOH |
| ▪ 2.8 - 4.17 V | ▪ 2.8 - 4.18 V | ▪ 2.8 - 4.2 V | ▪ 3.0 - 4.2 V |
| ▪ 0.1C / 0.33C | ▪ 0.1C / 0.33C | ▪ 0.1C / 0.5C | ▪ 0.1C / 0.5C |
| ▪ ~1.05 V after prelit. | ▪ ~1 V after prelit. | ▪ ~0.95 V after prelit. | ▪ ~0.7 V after prelit. |

*Gen2 = 1.2M LiPF$_6$ in EC:EMC 3:7 wt:wt

## Section S5. Additional supporting figures

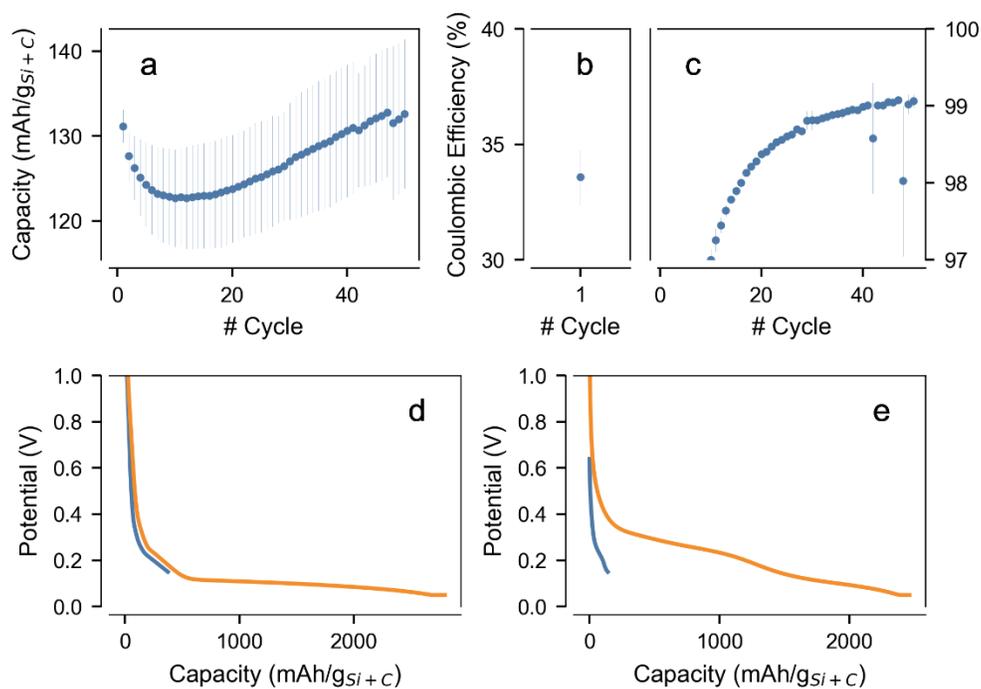

Figure S12. Data from cycling a c-Si electrode vs. Li metal in between 150 and 700 mV. a) capacity; b,c) coulombic efficiency; lithiation profiles in d) cycle 1; and e) cycle 3. In panels *d* and *e* the data in *orange* is from a cell lithiated to 50 mV that is provided as reference.

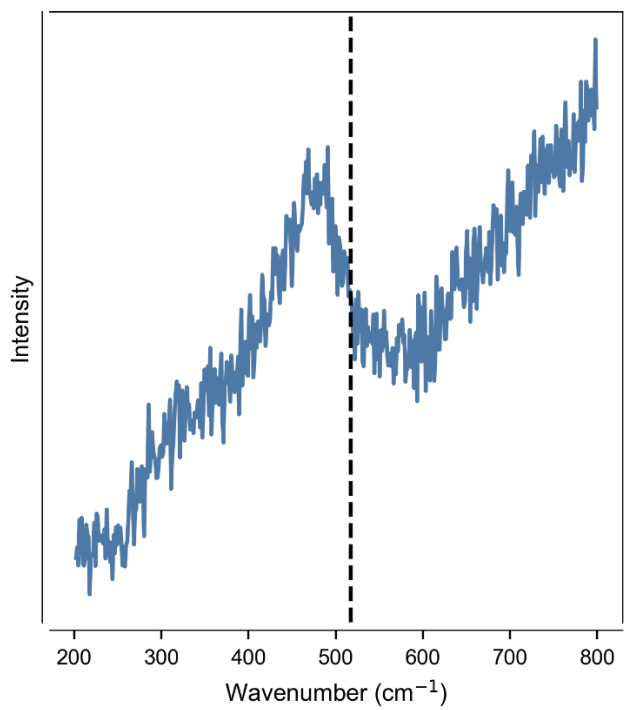

Figure S13. Example of Raman spectra for the SiO$_x$ material used in several of the cells shown in Figure 8. The dashed line indicates the position where the sharp band for crystalline Si is expected. Only amorphous domains were detected for this material.

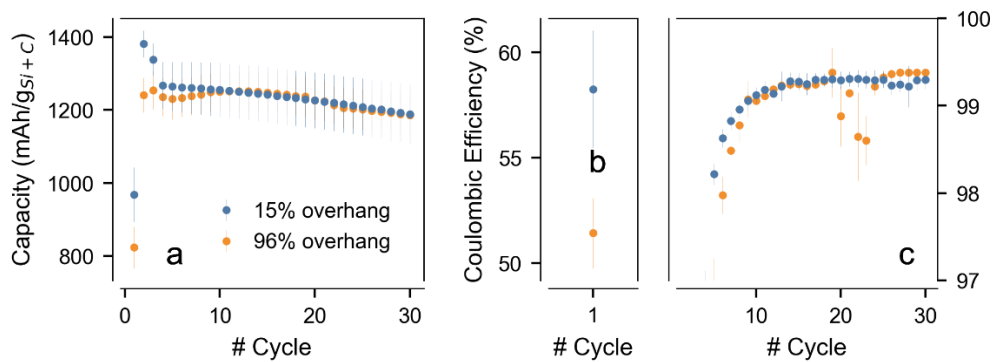

Figure S14. a) capacity and b,c) coulombic efficiency measured during cycling of NMC811 vs. a-Si cells at a slow C/10 rate. The negative electrode comprised 80wt% of nano-silicon (~200 nm, Paraclete Energy), 10wt% Timcal C45 carbon and 10wt% LiPAA binder, with a 1.10 mg/cm$^2$, a calendered coating of 10 µm and 47.3% porosity. The positive electrode is the same used in c-Si cells (see main manuscript). Cells were cycled in between 3.67 and 2.9 V, maintaining to NE within ~50 and ~600 mV vs. Li/Li$^+$.

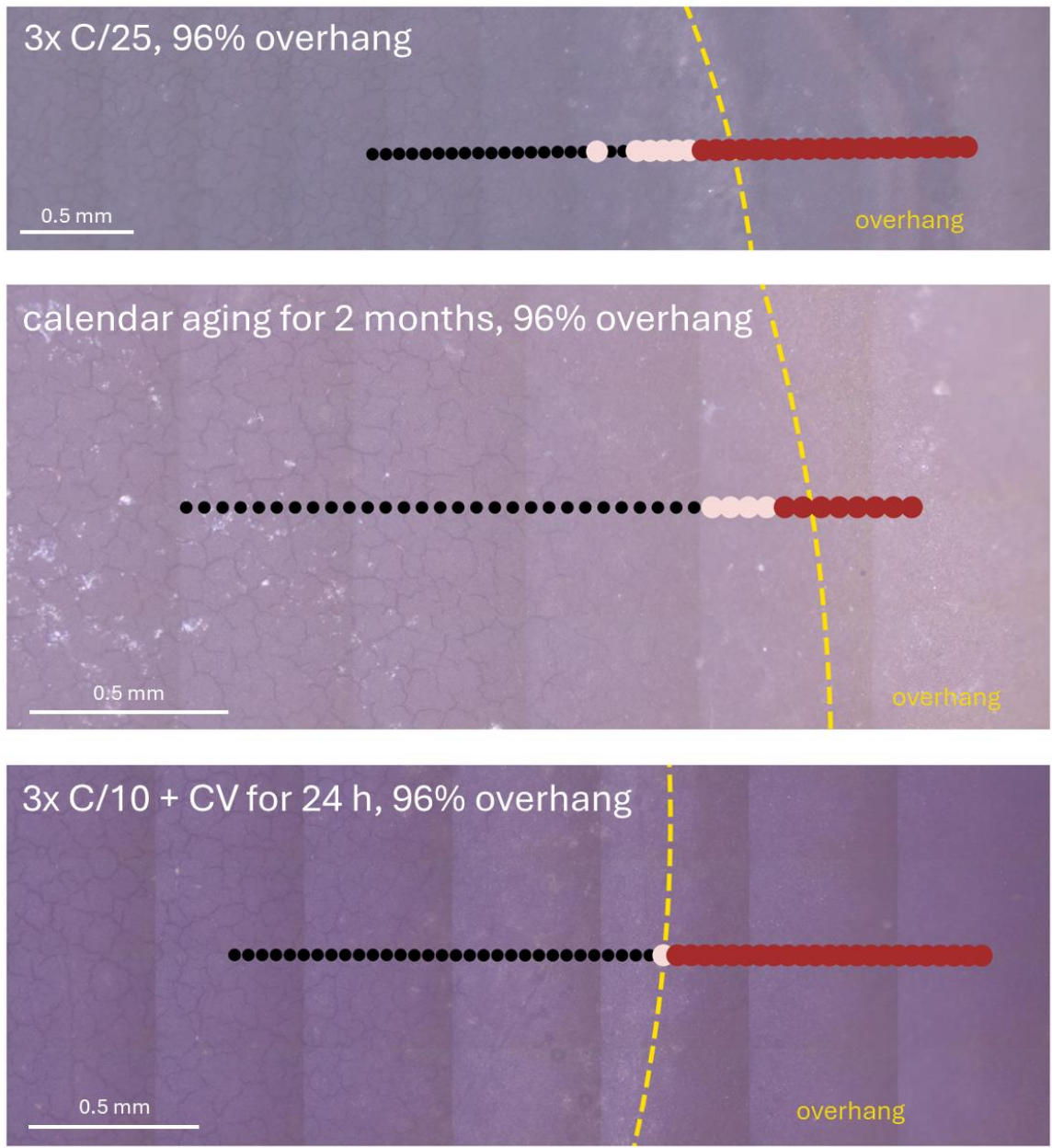

Figure S15. Example of optical images showing cracks in the electrode coating within the inner portions of the active area. The testing conditions in each case are specified at the top left of the images.

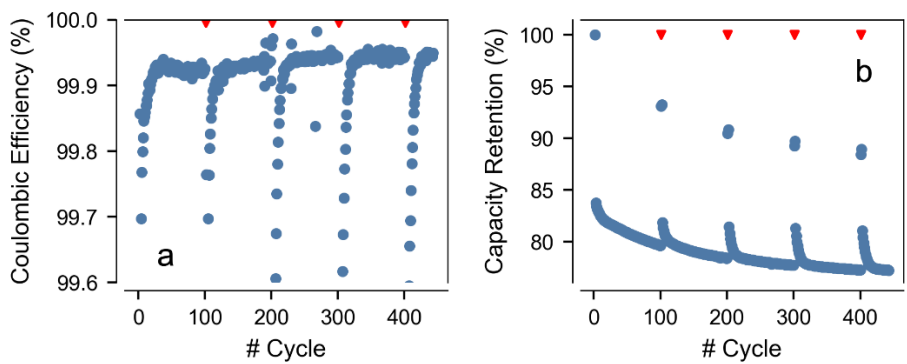

Figure S16. a) coulombic efficiency and b) capacity retention for a cell containing 80wt% Gr and 5wt% c-Si. The coulombic efficiency shows the trailing that is characteristic of overhang utilization. The position of RPTs is indicated by the red triangles at the top of the image. The anode was initially prelithiated electrochemically vs. Li metal, with a final delithiation to 500 mV.

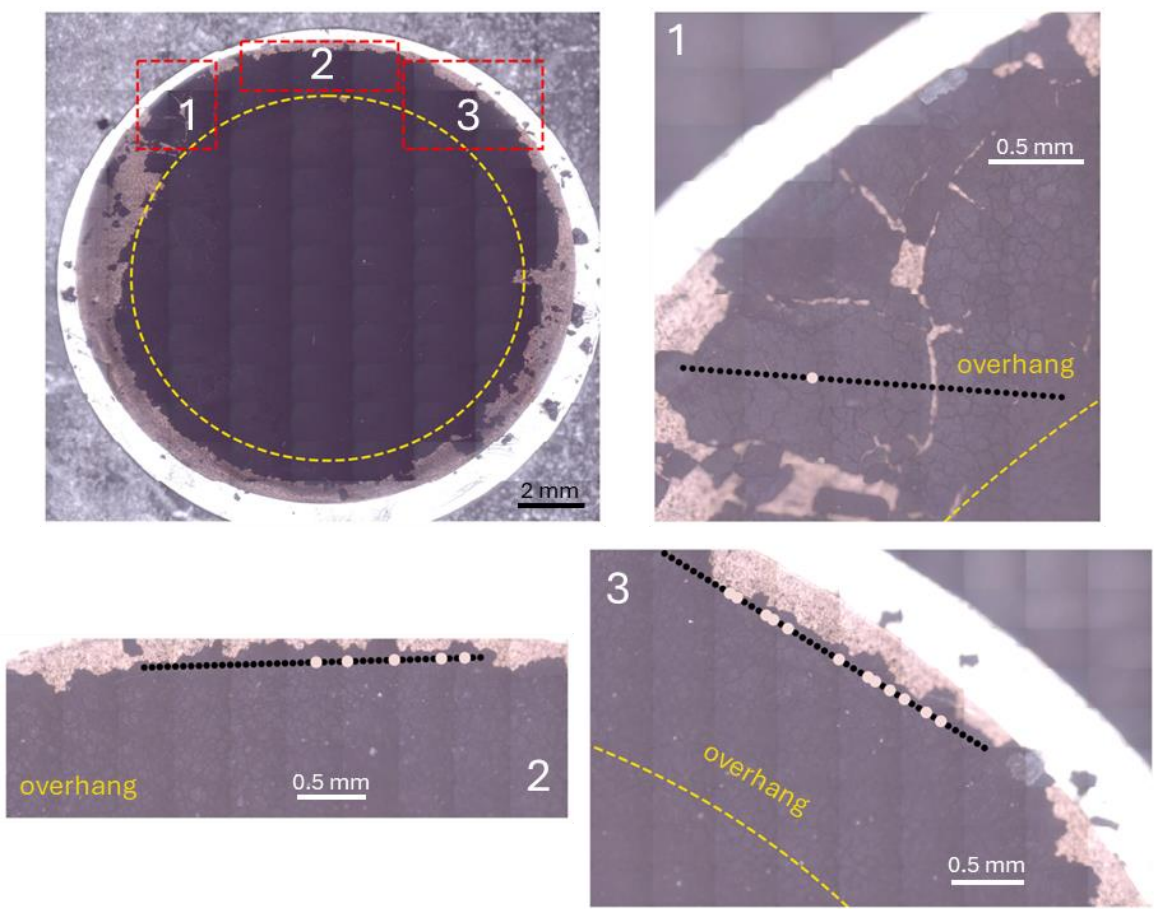

Figure S17. Raman studies with a c-Si electrode extracted from a full-cell with 38% overhang (11 mm Li metal vs. 14 mm c-Si) after 3x C/10 cycles, with a voltage hold at 50 mV until the current dropped to C/20. No c-Si bands were detected, indicating that the overhang was fully utilized. Also note the mud cracks being present at the overhang, providing additional evidence for extensive utilization. The Li chip was compressed prior to cell assembly to avoid changes to electrode size during crimping. Delamination occurred during cell disassembly.

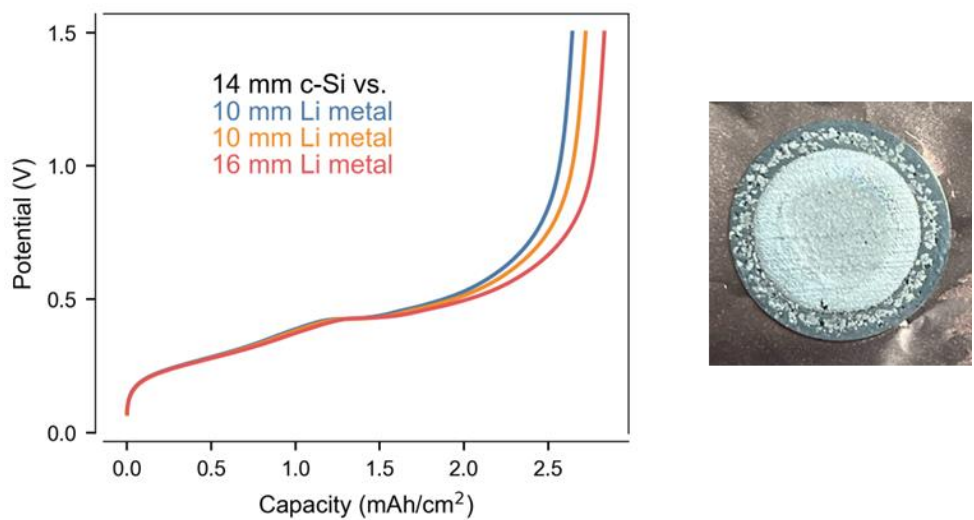

Figure S18. Delithiation capacity for c-Si vs. Li metal cells with and without an overhang of the Si electrode (*left*). Performance is similar for all cells, indicating that overhang can be promptly engaged. The *right* panel shows a photo of the Li metal extracted from a cell with a c-Si overhang, showing that Li$^+$ incoming from the overhang will tend to plate onto the steel spacer instead of at the Li disc.